\documentclass[prl,twocolumn,showpacs,superscriptaddress]{revtex4-1}

\usepackage{graphicx}
\usepackage{dcolumn}

\usepackage{eucal}
\usepackage[dvips]{epsfig}

\usepackage{amsmath,amsthm,amssymb}
\usepackage{bm}

\usepackage[dvipsnames]{xcolor}
\usepackage{verbatim}%Allows comments
\usepackage{float}

\newcommand{\be}{\begin{eqnarray}}
\newcommand{\ee}{\end{eqnarray}}
\newcommand{\bse}{\begin{subequations}}
\newcommand{\ese}{\end{subequations}}

\newcommand{\benn}{\begin{eqnarray*}}
\newcommand{\eenn}{\end{eqnarray*}}

\newcommand{\bnum}{\begin{enumerate}}
\newcommand{\enum}{\end{enumerate}}

\newcommand{\bit}{\begin{itemize}}
\newcommand{\eit}{\end{itemize}}

\newcommand{\bc}{\begin{cases}}
\newcommand{\ec}{\end{cases}}

\newcommand{\bpm}{\begin{pmatrix}}
\newcommand{\epm}{\end{pmatrix}}

\newcommand{\bvm}{\begin{vmatrix}}
\newcommand{\evm}{\end{vmatrix}}

\newcommand{\matx}[1]{{\bm{\mathsf #1}}}

\newcommand{\mbf}{\mathbf}
\newcommand{\mbb}{\mathbb}
\newcommand{\mcal}{\mathcal}

\newcommand{\ga}{\alpha}
\newcommand{\gb}{\beta}

\newcommand{\gd}{\delta}
\newcommand{\eps}{\epsilon}%\eps schon vergeben
\newcommand{\gf}{\phi}
\newcommand{\gl}{\lambda}

\newcommand{\gr}{\rho}
\newcommand{\gs}{\sigma}

\newcommand{\gvr}{\varrho}

\newcommand{\p}{\partial}
\newcommand{\f}{\frac}

\newcommand{\iy}{\infty}

\begin{document}
\title{Mode selection in compressible active flow networks}

\author{Aden Forrow}
%\email{aforrow@mit.edu}
\affiliation{Department of Mathematics, Massachusetts Institute of Technology, 77 Massachusetts Avenue, Cambridge MA 02139-4307, U.S.A.}
\author{Francis G. Woodhouse}
\affiliation{Department of Applied Mathematics and Theoretical Physics, Centre for Mathematical Sciences, University of Cambridge, Wilberforce Road, Cambridge CB3 0WA, U.K.}
\author{J\"orn Dunkel}
\email{dunkel@mit.edu}
\affiliation{Department of Mathematics, Massachusetts Institute of Technology, 77 Massachusetts Avenue, Cambridge MA 02139-4307, U.S.A.}

\date{\today}
\begin{abstract}
 Coherent, large scale dynamics in many nonequilibrium physical, biological, or information transport networks are driven by small-scale local energy input. Here, we introduce and explore an analytically tractable nonlinear model for compressible active flow networks. In contrast to thermally-driven systems, we find that active friction selects discrete states with a limited number of oscillation modes activated at distinct fixed amplitudes. Using perturbation theory, we systematically predict the stationary states of noisy networks and find  good agreement with a Bayesian state estimation based on a hidden Markov model applied to simulated time series data. Our results suggest that the macroscopic response of active network structures, from actomyosin force networks to cytoplasmic flows, can be dominated by a significantly reduced number of modes, in contrast to energy equipartition in thermal equilibrium. The model is also well-suited to study topological sound modes and spectral band gaps in  active matter.
\end{abstract}


\maketitle
%%%%%%%%%%%%%%%%%%%%%%%%%%%%%%%%%%%%%%%

Active networks constitute an important class of nonequilibrium systems spanning a wide range of scales, from the intracellular cytoskeleton \cite{Broedersz2014,Ruprecht:2015aa} and amoeboid organisms~\cite{Takamatsu2001,Tero2010_Science,Alim2013,Alim2017,Bonifaci2012,Reid2016} to macroscopic transport networks~\cite{Coclite2005,Piccoli2006,PhysRevE.89.020801,Heaton2012_PRE}. Identifying generic self-organization principles~\cite{Nakao2010_NatPhys,2008Mikhailov,Souslov2016} that control the dynamics of these biological or artificial far-from-equilibrium systems remains one of the foremost challenges of modern statistical physics. Despite promising experimental~\cite{Tero2010_Science,Alim2013, Alim2017,Fakhri2014,Ronceray2015,Marbach2016}  and theoretical~\cite{Broedersz2011,Bonifaci2012,Paoluzzi2015,Bressan2014,Broedersz2014} advances over the past decade, it is not well understood how the interactions between local energy input, dissipation and network topology determine the coordinated global behaviors of cells~\cite{Fakhri2014}, plasmodia~\cite{Tero2010_Science,Alim2013,Alim2017} or tissues~\cite{2016Martin}.  Further progress requires analytically tractable models that help clarify the underlying nonequilibrium mode selection principles~\cite{2000EbErDuJe,2001DuEbErMa}.

\par
We inroduce here a generic model for active flows on a network, motivated by recent experimental studies of bacterial fluids~\cite{2016Wioland_RaceTracks, Paoluzzi2015} and ATP-driven microtubule suspensions~\cite{2017Wueaal} in microfluidic channel systems.  Building on Rayleigh's work~\cite{Rayleigh1894} on driven vibrations and the Toner-Tu model of flocking~\cite{Toner2005}, the theory accounts for network activity  through a nonlinear friction~\cite{Toner2005,1998Schweitzer,2012RoGe,2012Burada_PRE}.  We work in a fully compressible framework allowing accumulated matter at vertices to affect flow through network pressure gradients, generalizing previous work on incompressible pseudo-equilibrium active flow networks~\cite{Woodhouse2016,Woodhouse2017}, as suited to the many biological systems exhibiting flexible network geometry~\cite{Tero2010_Science,Alim2013,Alim2017} or variations in the density of active components \cite{Souslov2016}.
Although inherently nonlinear, the model can be systematically analyzed through perturbation theory. Such analysis shows how slow global dynamics emerge naturally from the fast local dynamics, enabling prediction of the typical states in large noisy networks; these states have significantly fewer active modes than for energy equipartition~\cite{Khinchin} in thermal equilibrium. More broadly, our model provides an accessible framework for investigating generic physical phenomena in active systems, including topologically-protected sound modes~\cite{Souslov2016} and the influence of spectral band gaps (SM~\cite{SM}).

%%%%%%%%%%%%%%%%%%%%%%%%%%%
\begin{figure}[ht]
\includegraphics[width=0.82\linewidth,]{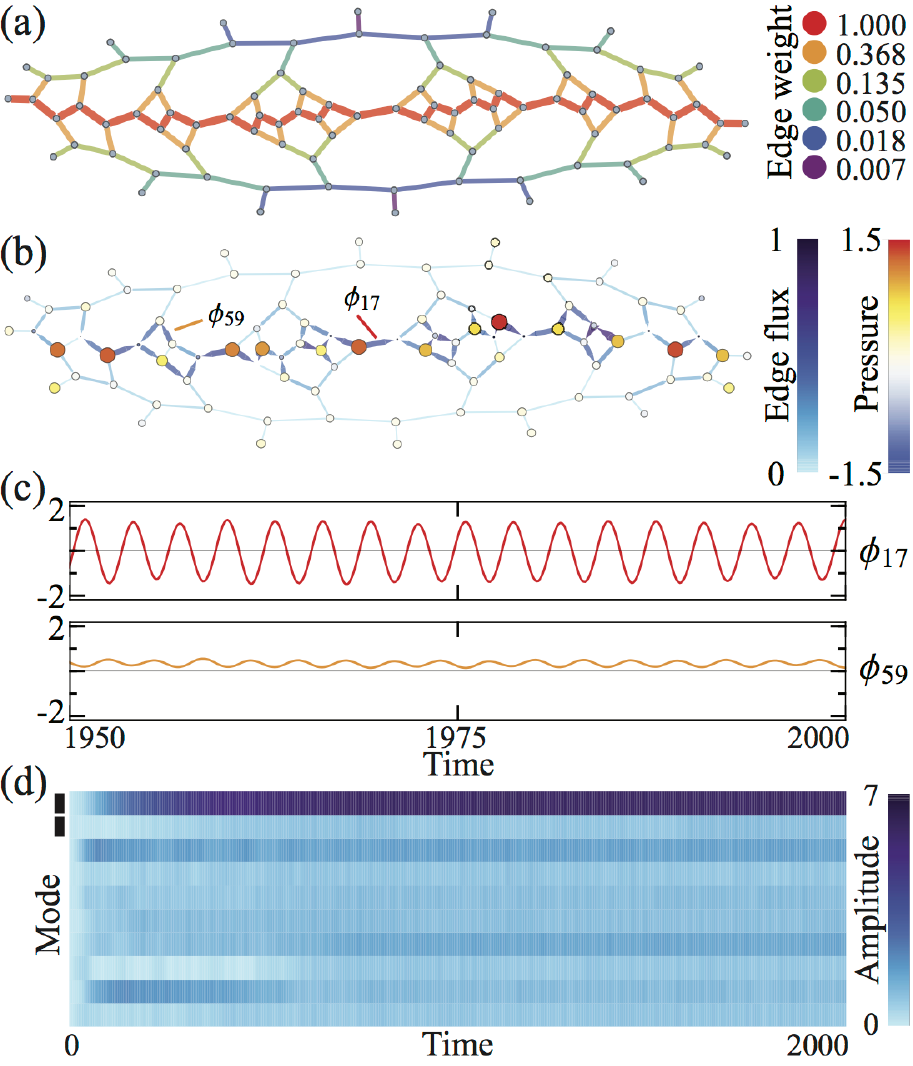}
\caption{
Activity can select a single dominant oscillation mode on hierarchically weighted networks.
(a) The edges in the graph simulated in (b) and (c) are given weights decreasing exponentially with their distance from the central red path.
(b) Oscillations in pressure and flux develop primarily along the central high-weight path (Movie 1).
(c) Edge fluxes $\phi_e$ settle into steady synchronized oscillations as exemplified for two edges indicated in (b), one on ($\phi_{17}$) and one off ($\phi_{59}$) the path.
(d) Plotting the time-dependent amplitude of each analytically-determined flow eigenmode confirms selection of a single oscillatory mode. The ten modes with the highest average amplitude in this simulation run are pictured;  the marked top two rows are oscillatory modes, while the remaining rows are cyclic modes. See Fig.~S6 for all modes.
Simulation parameters were $\epsilon = 0.1$, $\mu = 1$, and $D = 10^{-4}$.
} 
\label{f:hierarchical_cubic}
\end{figure}
%%%%%%%%%%%%%%%%%%%%%%%%%%%

We consider activity-driven mass flow on an arbitrarily-oriented graph $\mbb{G}=(\mcal{V},\mcal{E})$ with  $V=|\mcal{V}|$  vertices and $E=|\mcal{E}|$ edges. The elements of the $V\times E$ gradient (incidence) matrix $\nabla$ are $\nabla_{ve}=-1$ if edge $e$ is oriented outwards from vertex~$v$,  $\nabla_{ve}=+1$  if $e$ is oriented inwards into~$v$, and $\nabla_{ve}=0$ otherwise. The dynamical state variables are the deviations from the mean mass $\bar{\gvr}=M/V$ on the nodes, $(\gvr_1(t),\ldots,\gvr_V(t))$, and the mass fluxes on the edges, $(\phi_1(t),\ldots, \phi_E(t))$, governed by the non-dimensionalized (SM~\cite{SM}) transport equations
\bse
\label{e:active_model_resc}
\be
\label{e:active_model_resc-a}
\dot{\gvr}_v &=&  \sum_e\nabla_{ve} \phi_e,
\\
\label{e:active_model_resc-b}
\dot{\phi}_e &=&-  \sum_v \nabla^\top_{ev}\gvr_v + \eps \frac{\mu -\phi_e^2}{1+\phi_e^2}\, \phi_e
+ \sqrt{2 D}  \xi_e(t), \qquad
\ee
\ese
where $\xi_e(t)$ is standard Gaussian white noise. Equation~\eqref{e:active_model_resc-a} ensures mass conservation. The first term on the r.h.s. of Eq.~\eqref{e:active_model_resc-b} represents the gradient of an ideal gas-type node pressure $p_v\propto \gvr_v$, corresponding to the leading term in a virial expansion; the second term is a Toner-Tu type (SM~\cite{SM}) active friction force derived from a depot model~\cite{1998Schweitzer,2011Pawel_Chaos} with coupling $\eps>0$ and active--passive control parameter $\mu$, which drives the edge fluxes $\phi_e$ towards preferred values $\pm\sqrt{\mu}$ when $\mu > 0$. Many networks have non-uniform edge and vertex weights, which can be incorporated into equations of identical form to Eqs.~\eqref{e:active_model_resc} with appropriate rescaling of $\gvr$, $\phi$, and $\nabla$ (SM~\cite{SM}).

\par
Active flow networks described by Eqs.~\eqref{e:active_model_resc} exhibit rich oscillatory transport behavior, including the mode selection illustrated in Movie~1 and Fig.~\ref{f:hierarchical_cubic} for a hierarchically-weighted network with vertex degrees at most 3 as is typical of \textit{Physarum polycephalum} \cite{Baumgarten2010}. When this network is initialized with zero pressure variation and flux, it typically settles into a quasi-steady state with a single dominant oscillation frequency on the highest-weight path.  This is a manifestation of the fact that single-frequency selection is the norm on actively driven path graphs, as we shall show analytically below.  
\par
Generally, the features of the steady-state attractor will be determined by the topology of the subgraph of high-weight edges, which may be much sparser than the original network.
For this reason, as well as for ease of analysis and illustration, we will henceforth assume $\mbb{G}$ to be a tree, as realized in certain peripheral sensory neurons~\cite{2017Shura}, though in general the full model in Eqs.~\eqref{e:active_model_resc} is not restricted to any particular class of graph. The behaviors observed on trees can be extended to denser graphs by choosing appropriate edge weights.

The complex active flow dynamics encoded by Eqs.~\eqref{e:active_model_resc} can be understood analytically by considering the basis of oscillation modes of the network, as we illustrate now in the fully deterministic case ($D=0$).
To progress, we adopt a Rayleigh~\cite{Rayleigh1894} approximation $\eps (\mu-\phi_e^2) \phi_e$ for the active friction (SM~\cite{SM}).
Now, expand the pressure $\gvr_v = \sum_{n=1}^{E} r_n(t) \gvr_{vn}$ and flux $\phi_e = \sum_{n=1}^{E} f_n(t) \phi_{en}$ in the right and left singular vectors $\bm{\gvr}_n = (\gvr_{vn})$ and $\bm{\phi}_n = (\phi_{en})$ of~$\nabla^\top$ corresponding to the $E = V-1$ non-zero singular values $\lambda_n$.
(On a tree, there is a single zero eigenvalue of $\nabla\nabla^\top$ yielding an additional right singular vector for the pressure, but this corresponds to a constant mass shift and so can be safely neglected.)
Defining mode amplitudes $A_n^2 = r_n^2 +f_n^2$, the network energy then takes the simple form $H= \f{1}{2}\sum_n \lambda_n^2 A^2_n$
(SM~\cite{SM}).  
When $\eps$ is small there are two distinct timescales, namely the fast oscillation timescale $t$ and the slow friction timescale $\tau = \eps t$, which we separate in the perturbation ansatz $r_n=\sum_{\gs=0}^\infty \eps^\gs r_{\gs n}$ and $f_n=\sum_{\gs=0}^\infty \eps^\gs f_{\gs n}$~\cite{Strogatz2015}. Active friction does not contribute at lowest order, so the $O(1)$  contribution to each mode $(r_n, f_n)$ is an uncoupled harmonic oscillator $ r_{0n}(t) = A_{0n}(\tau) \cos[\gl_n t - \delta_n(\tau)]$ and $
f_{0n}(t) =-A_{0n} (\tau) \sin [\gl_n t - \delta_n(\tau)]$ with $t$-independent amplitude $A_{0n}$ and phase $\delta_n$  (SM~\cite{SM}).

%%%%%%%%%%%%%%%%%%%%%%%%
\begin{figure*}[t]
\includegraphics[width=0.825\linewidth]{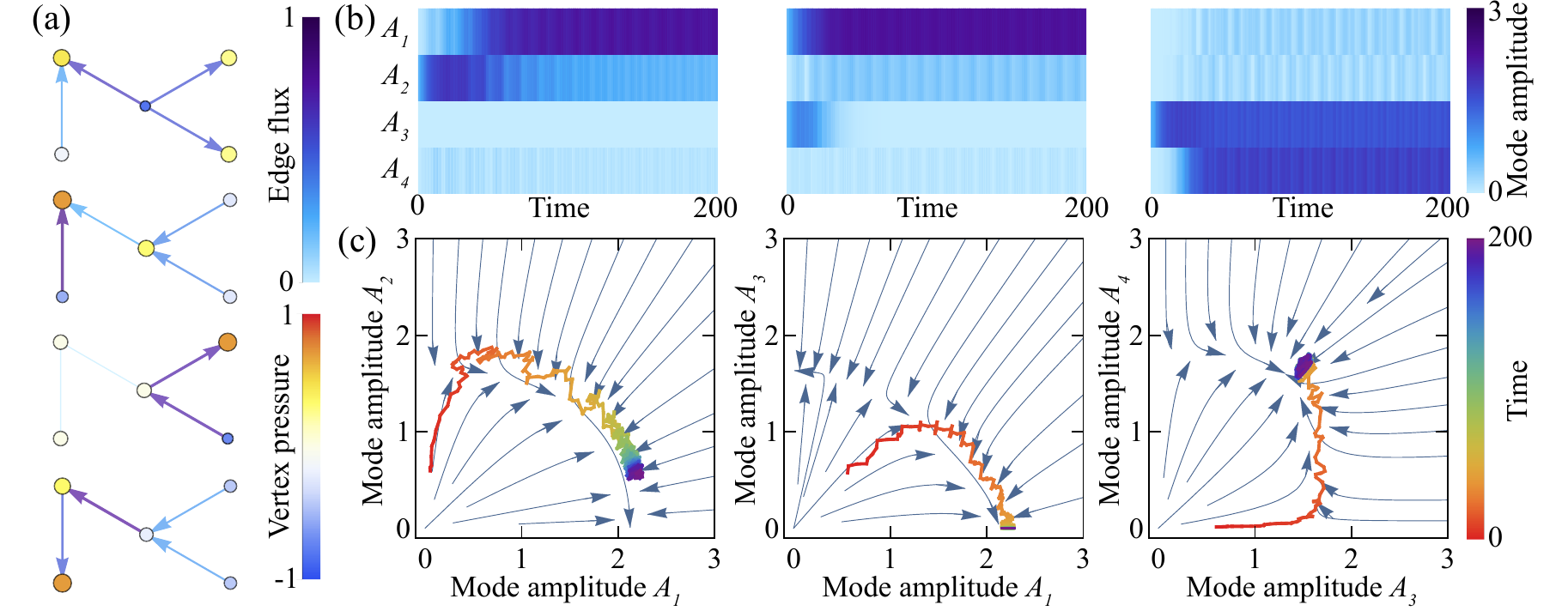}
\caption{First order perturbation theory accurately predicts the stable states on small trees. 
(a) A five vertex tree possessing four nontrivial modes, as illustrated. 
(b) On the tree in (a), mode amplitudes settle into one of two stable stationary states, as seen in simulations for three different initial conditions. Modes are ordered by frequency from high (top) to low (bottom).
(c) Simulated mode trajectories (rainbow) in (b) match analytic predictions (blue streamlines) in the subspaces of activated modes. There are three possible arrangements of nonzero critical points in each 2D subspace: a saddle point on one axis and a stable node on the other axis (left), a stable node on each axis and a saddle point in the middle (center), or a saddle point on each axis and a stable node in the middle (right; Movie 2). Higher order effects cause both the convergence to a point with $A_2 >0$ in the left and middle plots and the oscillations in the trajectories.  Parameters used are $\eps = 0.5$, $\mu = 1$, $D=0$. 
}
\label{f:t5}
\end{figure*}
%%%%%%%%%%%%%%%%%%%%%%%%%

\par
The influence of activity becomes apparent at first order in $\epsilon$, introducing couplings between mode amplitudes
whose dynamics encode the state selection behavior of the active network.
Requiring that the $O(\epsilon)$ amplitudes $r_{1n}$ and $f_{1n}$ remain small relative to the leading terms implies that the secular (unbounded) terms in the first order equations must vanish~\cite{Strogatz2015}. Assuming negligible mode degeneracies, the slow dynamics of the $O(1)$ mode amplitudes $A_{0n}(\tau)$ are found to obey~(SM~\cite{SM}) 
\be
\f{d (A_{0n}^2) }{d\tau}=  \left(\mu - \sum_{k=1}^E P_{nk} A_{0k}^2\right)\, A_{0n}^2,
\label{e:amplitude_dynamics}
\ee
where the overlap matrix
$P_{nk} = \tfrac{3}{2} (1 - \tfrac{1}{2}\gd_{nk})\sum_e \phi_{en}^2 \phi_{ek}^2$
encodes the network topology.
Fixed points of Eq.~\eqref{e:amplitude_dynamics} can then be found
by choosing a subset of the $A_{0n}$ to be zero and solving $\sum_{k=1}^E P_{nk} A_{0k}^2 = \mu$ for $A_{0n}^2$ over the remaining non-zero modes. If all the non-zero solutions for $A_{0n}^2$ are positive, then there is a stationary point with those modes activated~(SM~\cite{SM}).

Activity-driven fixed points with exactly one mode active always exist. If only mode $p$ is active at leading order, then $A_{0n} = \sqrt{\mu / P_{pp}} \,\gd_{np}$ is a fixed point of Eq.~\eqref{e:amplitude_dynamics}.
These amplitudes,
which closely match both those calculated with the full unapproximated active friction force and those from averages computed over fully nonlinear simulations~(SM~\cite{SM}), show that as  $\mu$ crosses $0$ there is a supercritical Hopf bifurcation with $A_{0n} \sim \sqrt{\mu}$.
However, the stability of such a single-mode state depends on topology: our 
simulations suggest that activity always selects exactly one oscillation mode in simple path graphs, whereas single-mode states are typically unstable in networks with complex topologies. We can use this observation to model more complex active networks with single mode selection by appropriately weighting the edges: if the edge weights for a path are large enough compared to the weights elsewhere in the network, the path behavior dominates (Fig.~\ref{f:hierarchical_cubic}).

Insight into stability is provided by the case with up to two modes active.
Writing $A_{0n} = A_{0p} \delta_{np} + A_{0q} \delta_{nq}$, Eq.~\eqref{e:amplitude_dynamics} yields
\be
{d(A_{0p}^2)}/{d\tau} = (\mu - P_{pp}A_{0p}^2 - P_{pq}A_{0q}^2)\, A_{0p}^2, 
\label{e:two_mode_amplitude}
\ee
and symmetrically for $A_{0q}^2$.
Depending on the topology-encoding overlap coefficients~$P_{nk}$, this gives up to four fixed points: the zero state  $A_{0p} = A_{0q} = 0$, which is always linearly unstable; the single-mode state $(A_{0p}, A_{0q}) = (\sqrt{\mu /P_{pp}}, 0)$, which is stable
if $P_{pq} > P_{pp}$ and a saddle if not, plus analogously for $(0,\sqrt{\mu /P_{qq}})$; and, potentially, a mixed state $(A_{0p}^*, A_{0q}^*)$ where $A_{0p}^*= \sqrt{ \mu \left({P_{qq} -  P_{pq}})\right/\left({ P_{pp} P_{qq}- P_{pq}^2 }\right)}$  with $A_{0q}^*$ defined symmetrically.
When it exists, the mixed state is either stable (if $P_{pq}^2 < P_{pp}P_{qq}$) or a saddle  (if $P_{pq}^2 > P_{pp}P_{qq}$), but if one of the single-mode states is stable and one is unstable, then one of $A_{0p}^*$ and $A_{0q}^*$ is imaginary and there is no mixed state.
Hence, we have three possible scenarios (Fig.~\ref{f:t5}): one stable single mode and the other a saddle with no mixed state (Fig.~\ref{f:t5}b,c; left); two stable single-mode states with a mixed saddle in-between (Fig.~\ref{f:t5}b,c; center); and two single-mode saddles with a stable mixed state in-between (Fig.~\ref{f:t5}b,c; right).
These predictions match simulations quantitatively even for relatively large $\eps$ beyond the  small-$\epsilon$ perturbation regime (Fig.~\ref{f:t5}).
In fact, simulations show the same qualitative behavior for $\eps = 2$, suggesting  perturbation analysis remains predictive at high activity.

%%%%%%%%%%%%%%%%%%%%%%%%
\begin{figure*}[ht]
\includegraphics[width=0.825\linewidth]{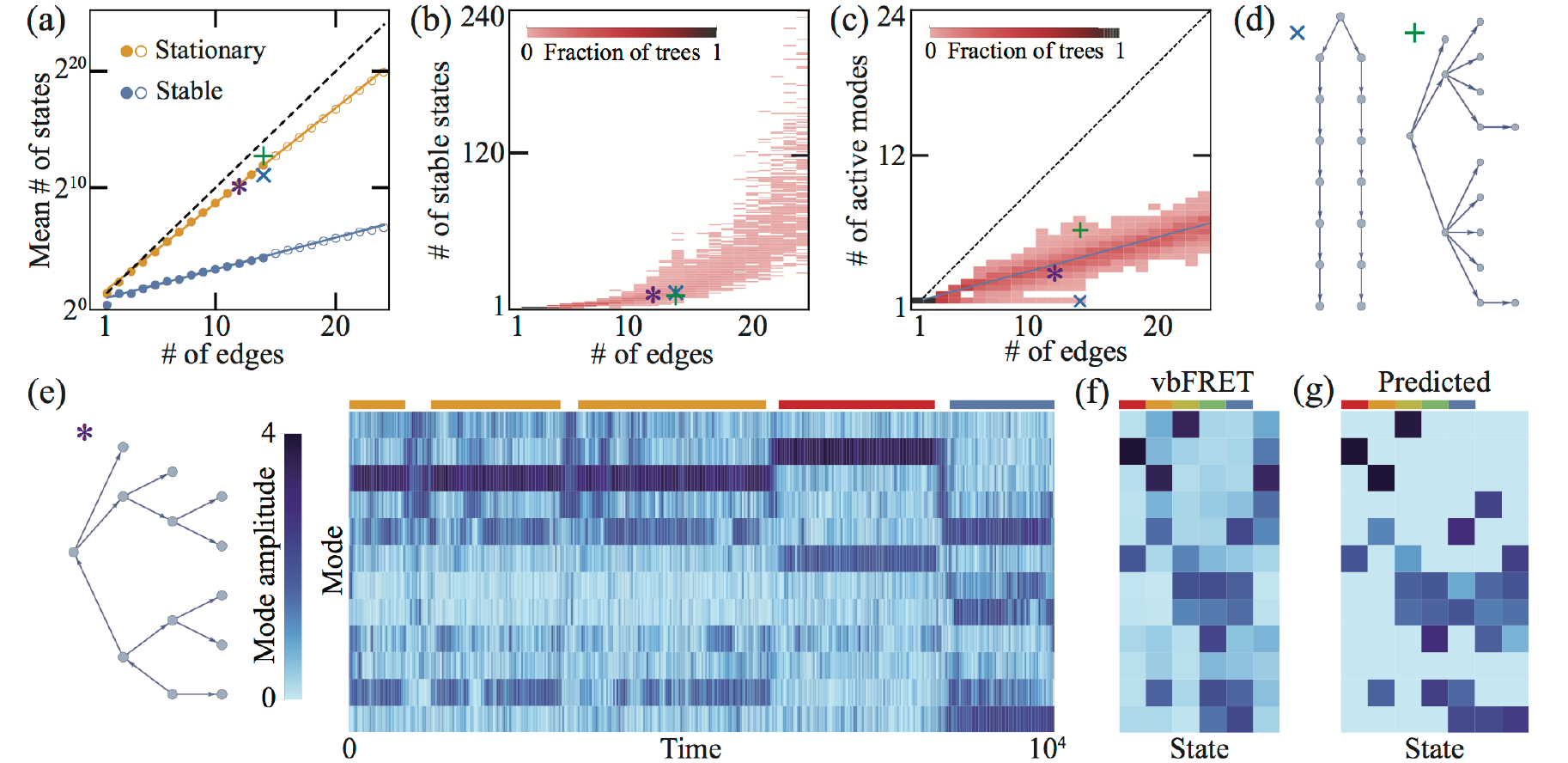}
\caption{States on larger trees possess surprisingly few active modes, which can be inferred from  time series with non-zero noise.
(a)~The mean number of stationary states of Eq.~\eqref{e:amplitude_dynamics} grows exponentially with edges $E$ as $1.77^E \approx (2^E)^{4/5}$  (solid orange line), close to the upper bound of $2^E$ states (dashed black line), while the mean number of stable states grows as $1.21^E \approx (2^E)^{1/4}$ (solid blue line). We counted states on all nonisomorphic trees with $E\le  14$ edges (filled circles) and on a random sample of $\sim 175$ trees per point for $15 \le E \le 24$ (open circles). Averages are over trees with a fixed number of edges. 
(b)~As $E$ increases, both the mean and the variance of the distribution of trees with each number of stable states increase rapidly.
(c)~Distribution of the average number of modes active in a stable state. The mean over trees scales like $0.26 E\approx E/4$ (solid line), significantly below $E/2$ expected if modes were selected randomly.
(d)~Two example trees indicated in (a-c) by the corresponding colored symbols. Stable states on paths ($\times$) always only activate one mode; complex trees ($+$) have more modes active.
(e)~Noisy networks ($D>0$) transition stochastically between stable states, exemplified by an amplitude-time trace for the tree shown. 
Modes are ordered by frequency from high (top) to low (bottom). Simulation parameters are $\eps = 0.5$, $\mu = 1$, $D = 5 \times 10^{-3}$. 
(f)~States found by vbFRET from simulations on the tree in (e)~(SM~\cite{SM}). The second, first, and fifth columns are states seen in (e), indicated by the colored bars above.
(g)~States predicted by Eq.~\eqref{e:amplitude_dynamics} for the tree in (e). The first five states in (f) match those in (g); the sixth column in (f) is likely a transient combination of analytically stable states. 
}
\label{f:states}
\end{figure*}
%%%%%%%%%%%%%%%%%%%%%%%%

This two-mode analysis yields a simple topological heuristic for the stability of single-mode states. Since $|\bm{\phi}_p|$ = 1, $P_{pp}$ is small when $\bm{\phi}_p$ is spread over many edges and large when $\bm{\phi}_p$ is localized to a few edges. If $\bm{\phi}_q$ is localized to the same edges as $\bm{\phi}_p$, $P_{pq}$ will also be large and mode $p$ will be stable to perturbations in mode $q$. However, if $\bm{\phi}_q$ is localized to a disjoint set of edges,
$P_{pq}$ will be a scaled inner product of near-orthogonal vectors $(\phi_{ep}^2)$ and $(\phi_{eq}^2)$ and will be small. Thus localized modes will be unstable to modes in other regions, while conversely if a mode is to be stable alone then it will be spread out across the entire network. Therefore, a stable combination of modes will possess significant flows on all edges of the network.

Biological systems exhibit vastly different macroscopic and microscopic time scales~\cite{Halatek2012,Kerr2006,Riedel-Kruse2007,Varma2008}. This phenomenon is present in our compressible active flow network, where higher-order nonlinear effects induce slow global time scales from faster small-scale dynamics.
When the zeroth-order amplitudes $A_{0n}$ are at a fixed point, the first-order corrections $r_{1n}$ and $f_{1n}$ are harmonic oscillators with natural frequency $\gl_n$ driven at linear combinations of the frequencies active at zeroth order (SM~\cite{SM}). For instance, if two modes $p$ and $q$ are active at zeroth order, the driving frequencies are $3\gl_p - k(\gl_p\pm\gl_q)$ for $k = 0, \ldots, 3$. This introduces new, slower timescales into the dynamics, including oscillations in the energy $H = \f{1}{2}\sum_n \gl_n^2 (r_n^2 + f_n^2)$ with frequency $\gl_p-\gl_q$. Their magnitude depends on the difference in frequency: slower oscillations, driven by modes with similar frequencies $\gl_p \approx \gl_q$, have higher amplitudes (SM~\cite{SM}, Fig.~S7).

The number of activated modes in an arbitrary compressible active network depends on intricate interactions between local activity and global flow configurations. The total number of available modes is equal to the number of edges $E$, meaning that, were each combination of modes to be a fixed point, a tree could have up to $2^E$ stationary states. To see how the true number of stationary and stable states depends on tree size, we performed an exhaustive numerical fixed point search of Eq.~\eqref{e:amplitude_dynamics} over a large sample of trees with $E\le  24$ (Fig.~\ref{f:states}a-d). The naive upper bound of $2^E$ suggests exponential growth of the mean number of steady states with edges $E$; this is indeed what we see, going as $\sim(2^E)^{4/5}$. However, though still exponential in $E$, the mean number of stable states is much smaller at $\sim (2^{E})^{1/4}$ (Fig.~\ref{f:states}a). Remarkably, these stable states have only $\sim E/4$ modes active on average (Fig.~\ref{f:states}c) in stark contrast to the activation of all $E$ modes under thermal equipartition~\cite{Khinchin}. Path-like topologies lead to even more dramatic reductions in the number of modes active (Fig.~\ref{f:states}c), suggesting that a biological system can further reduce the number of active modes through an optimal choice of topology; moreover, hierarchically tuned edge capacities as realized in \textit{Physarum}~\cite{Baumgarten2010,Alim2013,Alim2017} can further enhance mode selection even in non-tree topologies (Fig.~\ref{f:hierarchical_cubic}).

Real active transport networks will have some nonzero level of thermal or athermal  noise~\cite{lindner2004effects,1998GaHaJuMa,Hanggi1990_RevModPhys}. Provided the noise is not too large, it will render previously stable states now only metastable, with flow patterns exhibiting small fluctuations around these metastable states punctuated by noise-driven stochastic transitions between them~\cite{Hanggi1990_RevModPhys,Woodhouse2016}. Long-time simulations of Eqs.~\eqref{e:active_model_resc} with $D >0$ therefore offer an  independent numerical way to find stable fixed points of the amplitude dynamics. We use  vbFRET \cite{Bronson2009}, a variational Bayesian analysis of a continuous time hidden Markov model, to identify states from simulated time series. Almost all of the states discovered by vbFRET  match stable states predicted by Eq.~\eqref{e:amplitude_dynamics} even in the presence of non-negligible noise (Fig.~\ref{f:states}e-g), justifying the simplifications used in deriving Eq.~\eqref{e:amplitude_dynamics}. This also promises that Bayesian methods like vbFRET will function as reliable inference tools for experimental data from real-life active flow networks~\cite{Tero2010_Science,Alim2013,Marbach2016}.

Beyond active density oscillations~\cite{Paoluzzi2015}, the above theoretical framework can be used to probe the effects of topology on the physical properties of complex active systems. For instance, it was recently shown that continuum Toner--Tu systems in finite lattice confinement possess topologically protected edge-localized sound modes~\cite{Souslov2016}. Similar edge modes can be reproduced in our coarse-grained model through a simplified network representation of complex channel geometries (SM~\cite{SM} and  Movie~3). In addition, generalizing to allow different effective weights at vertices opens up band gaps, reflected in the excitation spectrum of spontaneous activity modes (SM~\cite{SM}). As we focus on  phenomenological  properties shared by many active systems, akin to the Toner--Tu approach~\cite{Toner2005}, the results and techniques presented here promise insights into the mode selection mechanisms governing a wide range of non-equilibrium  transport and force networks.

This work was supported by NSF Award CBET-1510768 (A.F. and J.D.), Trinity College, Cambridge~(F.G.W.), and an Alfred P. Sloan Research Fellowship~(J.D.). The authors thank Martin Zwierlein for stimulating discussions on band gaps.


%%%%%%%%%%%%%%%%%%%%%%%%%%%%%%%%%%%%%%%%%%
%%%%%%%%%%%%%%%%%%%%%%%%%%%%%%%%%%%%%%%%%%
%%%%%%%%%%%%%%%%%%%%%%%%%%%%%%%%%%%%%%%%%%
\appendix
\newpage
\null
\newpage

\begin{widetext}

\begin{center}
\large{\bf Supplemental material:
Mode selection in compressible active flow networks}
%\title{Supplemental material:
%Active flows on trees}
\vspace{5mm}

Aden Forrow,
Francis G. Woodhouse,
and J\"orn Dunkel

\end{center}

\renewcommand\figurename{Fig.~S$\!\!$}
\setcounter{figure}{0}

\section{Nondimensionalization of governing equations}

We can define the model in terms of the dimensional quantities $\hat \gvr_v$, $\hat \phi_e$, and $\hat t$; global dimensional parameters $\hat \epsilon$, $\hat \beta$, and $\hat D$; dimensionless edge conductances $\gamma_e$ and vertex volumes $m_v$; and a dimensionless global parameter $\mu$ and function $g$ as

\benn
\f {d {\hat \gvr}_v}{d \hat t} &=&  \sum_e \nabla_{ve} \hat \phi_e,
\\
\f {d{\hat \phi}_e}{d \hat t} &=&- \hat\gamma_e \sum_v  \nabla^\top_{ev} m_v ^{-1} \hat \gvr_v +  \hat \epsilon g\biggl(\mu, \f {\hat \phi_e}{\hat\beta\hat\gamma_e}\biggr)\hat \phi_e
+ \sqrt{2 {\hat D}} \hat \xi_e(\hat t).
\eenn

The scaling by conductance in the argument of $g$ is chosen to match the phenomenology observed in dense bacterial suspensions, where activity selects a characteristic velocity $\phi_e/\gamma_e$ and not a fixed flux $\phi_e$.
If we choose a conductance scale $\hat \gamma$ and volume scale $\hat m$ and insert the rescaled, nondimensional parameters
 \begin{align*}
  \gamma_e = \hat \gamma^{-1} \hat \gamma_e,
 \qquad
 m_v = \hat m^{-1} \hat m_v,
 \qquad
 \eps = \hat \gamma^{-\frac{1}{2}} \hat \eps,
 \qquad
 D_e =\hat \beta^{-2} \hat \gamma^{-\frac{1}{2}} \gamma_e^{-1}\hat D 
 \end{align*}
 and variables
 \begin{align*}
\gvr_v = m_v^{\frac{1}{2}} \hat\gamma^{\frac{1}{2}}\hat\beta^{-1}   \hat \gvr_v,
\qquad
 \phi_e = \gamma_e^{-\frac{1}{2}}\hat\beta^{-1}\hat \phi_e,
 \qquad
  t = \hat\gamma^{\frac{1}{2}}\hat t  ,
  \qquad
 \xi_e(t) = \hat\gamma^{-\frac{1}{4}}\hat \xi_e(\hat t),
 \end{align*} 
  we are left with 
\benn
 \f {d  \gvr_v}{d t} &=&   \sum_e m_v^{-1/2}\nabla_{ve} \gamma_e^{1/2} \phi_e,
\\
\f {d{ \phi}_e}{d t} &=&-  \sum_v \gamma_e^{1/2} \nabla^\top_{ev}m_v^{-1/2} \gvr_v + \epsilon g\biggl(\mu, \f { \phi_e}{\sqrt{\gamma_e}}\biggr)\phi_e
+ \sqrt{2  D_e} \xi_e( t).
\eenn
 
With constant conductances $\gamma_e = 1$ and volumes $m_v = 1$, we recover the model introduced in the main text, namely
\bse
\label{e:active_model_resc}
\be
\label{e:active_model_resc-a}
\f {d  \gvr_v}{d t} &=& \sum_e\nabla_{ve}  \phi_e,
\\
\label{e:active_model_resc-b}
\f {d{ \phi}_e}{dt} &=&-\sum_v \nabla^\top_{ev} \gvr_v + \eps g(\mu,\phi_e)\phi_e
+ \sqrt{2 D}  \xi_e(t) ,
\ee
\ese
with nonzero entries of the gradient matrix equal to $\pm 1$. 
All of our analysis applies equally well to the varying weights case: the only substantive change is replacing $\nabla_{ve}$ with the weighted gradient $\nabla^*_{ve} =  m_v^{-1/2}\nabla_{ve} \gamma_e^{1/2}$.

We can combine Eqs.~\eqref{e:active_model_resc-a} and \eqref{e:active_model_resc-b} into one second order equation for the pressure dynamics reading
\be
\ddot \gvr_v = \sum_e\nabla_{ve}  \left(-\sum_u \nabla_{eu}^\top \gvr_u + \eps g(\mu,\phi_e)\phi_e
+ \sqrt{2 D}  \xi_e(t) \right).
\label{e:rho_second_order}
 \ee
 In the absence of friction, when $g(\mu, \phi_e) = 0$, the dynamics are Hamiltonian with energy
 \be
 H =\f 1 2 \sum_{v,e,u} \gvr_v \nabla_{ve} \nabla_{eu}^\top \rho_u + \f 1 2 \sum_{e,v,f} \phi_e \nabla_{ev}^\top \nabla_{vf} \phi_f.
 \ee
The energy is particularly simple when written in the basis of singular vectors of $\nabla^\top$ with non-zero singular values, giving
\benn
H = \f 1 2 \sum_n \gl_n^2 \left( r_n^2 + f_n^2\right) \equiv \sum_n H_n.
\eenn

\section{Relation to physical flow systems}
We chose to explore a minimal model coupling local active energy input to network structure, rather than capture the details of any particular model system. Nevertheless, the key features of our model, namely mass conservation and a polynomial expansion of the active term, are generic enough to be straightforwardly adapted to a range of applications.

Mass conservation and pressure driven flow are likely to remain in any active flow model; the form of the active term may change in different contexts. In our case, staying close to examples of bacterial suspensions, we model activity as driving spontaneous flow on all edges. An alternative option, more closely related to shuttle streaming in networks, would be to apply an active force $f_v$ that compresses or expands each vertex and drives flow in or out, with modified dynamics
\bse
\label{e:physarum_model}
\benn
\label{e:physarum_model-a}
\f {d  \gvr_v}{d t} &=& \sum_e\nabla_{ve}  \phi_e,
\\
\label{e:physarum_model-b}
\f {d{ \phi}_e}{dt} &=&-\sum_v \nabla^\top_{ev}( \gvr_v + \eps f_v)
+ \sqrt{2 D}  \xi_e(t)  .
\eenn
\ese
The correct form of the active force depends on the microscopic details of the driving. Some generic features, however, will not depend on the exact form of $f_v$ and will be discoverable by choosing a simple function of local quantities ($\gvr_v, \dot \gvr_v$, etc.) as an approximate driving force.

The same method is used to derive the Toner-Tu equations for continuous active flows \cite{Toner2005};
our model can be understood as a discrete version of a special case of these equations.
If advective and diffusive terms are rendered negligible in favor of pressure-driven and activity-driven flow by geometric effects or otherwise, and we take only the linear term in the virial expansion of the active pressure, the general Toner--Tu model simplifies to
\bse
\benn
\frac{\partial \vec{v}}{\partial t}
= \ga \vec{v} - \gb |\vec{v}|^2 \vec{v} - \gs_1\vec{\nabla} (\rho - \rho_0)
 + \vec{f},
\qquad
\frac{\partial \gr}{\partial t} + \vec{\nabla} \cdot (\vec{v} \gr)  = 0.
\eenn
\ese
In a limit where deviations from the mean density are small, so $\gr = \gr_0 + \eta \gvr$ for some $\eta \ll 1$, we can further reduce to
\bse
\benn
\frac{\partial \vec{v}}{\partial t}
= \ga \vec{v} - \gb |\vec{v}|^2 \vec{v} - \eta\gs_1\vec{\nabla} \gvr
 + \vec{f},
\qquad
\eta \frac{\partial \gvr}{\partial t} + (\rho_0 +\eta\gvr)\vec{\nabla} \cdot \vec{v} +\eta \vec{v}\cdot \vec{\nabla} \gvr  = 0.
\eenn
\ese
Then on short time scales $\tau = t/\eta$, we have
\bse
\benn
\frac{\partial \vec{v}}{\partial \tau}
= \eta\ga \vec{v} - \eta\gb |\vec{v}|^2 \vec{v} - \eta^2\gs_1\vec{\nabla} \gvr
 + \eta\vec{f},
\qquad
 \frac{\partial \gvr}{\partial \tau} \approx -\rho_0\vec{\nabla} \cdot \vec{v},
\eenn
\ese
where we neglect terms that must be of order $\eta$: if the coefficients $\ga$, $\gb$, and $\gs_1$ are sufficiently large, their terms will remain relevant. The scaling of $\tau$ ensures that $t$ is small when $\tau$ is order one or smaller.  Discretizing the velocity and density fields as well as the noise $\vec{f}$ and replacing the continuous gradient with either $\nabla_{ev}^\top$ or $-\nabla_{ve}$ as appropriate yields Eqs.~\eqref{e:active_model_resc}.

\section{Compressibility}
Compressibility as included in our model is intended to describe changes in density or volume of the active component, not the underlying fluid. For example, variations in $\gvr$ may be interpreted as variations in the density of swimmers in a bacterial system or variations in the tube volume in \textit{Physarum polycephalum}. Such systems may be effectively compressible even though the solvent fluid (e.g. water) is incompressible.

In some cases, compressibility is the primary object of interest. For example, a recent preprint \cite{Souslov2016} discusses sound in active fluids in a network using a continuous wave equation derived from the Toner-Tu model. On top of a background flow taking the form of a lattice of counter-rotating cycles, they find modes confined to the edges of a Lieb lattice, which we can reproduce in our discretized setting (Fig.~S\ref{f:lieb} and Movie 3). In both their setting and ours, these edge modes decay over time without propagating into the bulk (cf. discussion in App. I.B of  Ref.~\cite{Souslov2016}).

We can recover an incompressible limit of our model by first extending it to include damping on the vertices:
\bse
\label{e:model_rho_damping}
\be
\label{e:model_rho_damping-a}
\f {d  \gvr_v}{d t} &=& \sum_e\nabla_{ve}  \phi_e - \eta \gvr_v,
\\
\label{e:model_rho_damping-b}
\f {d{ \phi}_e}{dt} &=&-\gamma\sum_v \nabla^\top_{ev} \gvr_v + \eps g(\mu,\phi_e)\phi_e
+ \sqrt{2 D}  \xi_e(t)  .
\ee
\ese
This paper examines the limit $\eta \rightarrow 0$ where total mass is exactly conserved. Previous work \cite{Woodhouse2016} has looked at the opposite limit, $\eta \rightarrow \infty$, where Eq.~\eqref{e:model_rho_damping-a} can only be balanced if $\gvr_v \rightarrow 0$ and
\benn
\gvr_v = \frac{1}{\eta} \sum_e\nabla_{ve}  \phi_e.
\eenn
Substituting this into Eq.~\eqref{e:model_rho_damping-b} gives
\benn
\f {d{ \phi}_e}{dt} &=&-\frac{\gamma}{\eta}\sum_v \nabla^\top_{ev} \nabla_{va} \phi_a + \eps g(\mu,\phi_e)\phi_e
+ \sqrt{2 D}  \xi_e(t). 
\eenn
With $g(\mu, \phi_e) = \phi_e^2 (1-\phi_e^2)$, this is equivalent to the model discussed in \cite{Woodhouse2016}. If $\gamma \rightarrow \infty$ so that $\gamma/\eta$ is constant, small deviations from incompressibility are allowed; if $\gamma/\eta \rightarrow \infty$, incompressibility is fully enforced. However, compressibility is a necessary ingredient for sound waves~\cite{Souslov2016}  and density oscillations~\cite{Paoluzzi2015}.

\begin{figure*}[t]
\includegraphics[width= 0.8\linewidth]{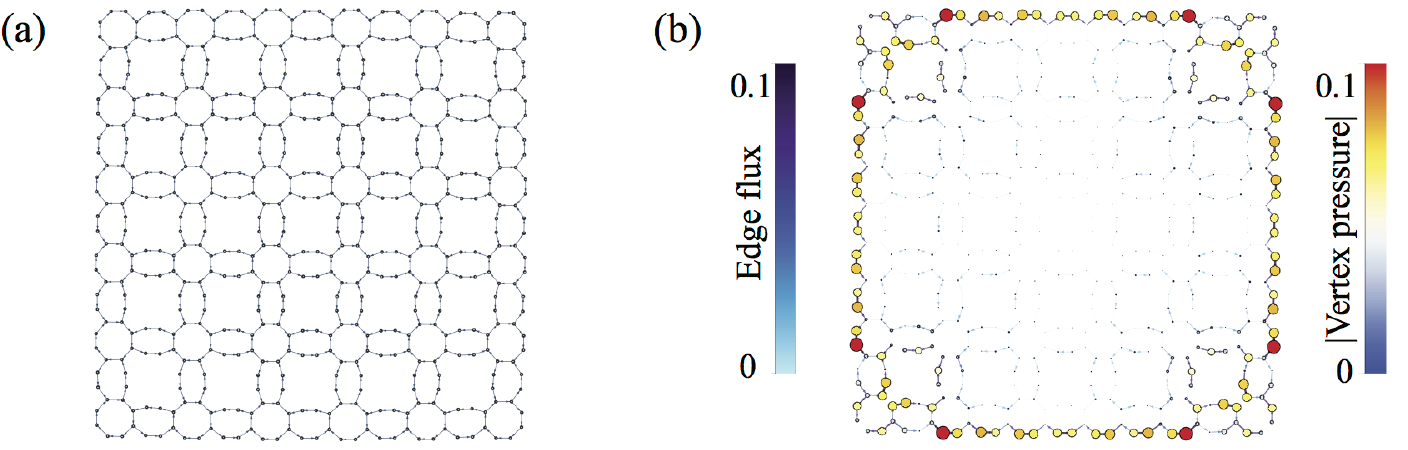}
\caption{Our active network model exhibits behavior similar to the topological edge modes of~\cite{Souslov2016}. 
(a) A discretized version of the Lieb lattice considered in \cite{Souslov2016}. Edges shared by adjacent 8-cycles have weight $\gamma_e =2$ to account for the additional width of the corresponding channels. The most stable flow on this network consists of a lattice of counter-rotating cycles, in which both the active friction term $g(\mu,\phi_e/\sqrt{\gamma_e})$ and the pressure variations $\gvr_v$ are everywhere zero.
(b) This lattice has modes confined to the edges of the domain, allowing sound waves to propagate and decay without scattering into the bulk (cf. discussion in App. I.B of  Ref.~\cite{Souslov2016}); one such mode is pictured.  
Simulations started in this mode as a perturbation to the most stable flow pattern do not cause density changes in the center (Movie 3). The network model allows study of such phenomena without resorting to full scale simulation of the flow patterns. }

\label{f:lieb}
\end{figure*}

\section{Rayleigh friction approximation}
While choosing the friction function to be~\cite{1998Schweitzer}
\benn
g(\mu, \phi_e) = \frac{\mu -\phi_e^2}{1+\phi_e^2} 
\eenn
has convenient theoretical properties, namely that it gives a passive constant friction coefficient $\eps$ for $\mu = -1$ and for $\phi \rightarrow \infty$, it is analytically difficult. To simplify the analysis, we approximate this $g(\mu, \phi_e)$ with a symmetric quadratic~\cite{Rayleigh1894}
\be
\hat g (\mu, \phi_e) = a - b \phi_e^2,
\label{e:quadratic_activity}
\ee
where $a = \mu$ and $b = 1$ are chosen so that $\hat g(\mu, 0) = g(\mu, 0)$ and $\hat g(\mu, \phi_e)$ has the same zeros as $g(\mu, \phi_e)$. This ensures that the two functions approximately match when they are both negative, that is, when activity is putting energy into the flow. The large difference between $g(\mu, \phi_e)$ and $\hat g(\mu, \phi_e)$ when the flux is large is less important, as the flow will be damped down in either case. The larger damping in $\hat g(\mu, \phi_e)$ does result in slightly lower steady amplitudes, both analytically and in simulations.

\section{Perturbation expansion}

 If $\eps$ is small, there will be two widely separated timescales: the fast oscillation timescale $t$ and the slow friction timescale $\tau = \eps t$. After writing $\gvr_v$ and $\phi_e$ in the mode basis, we can further expand in $\eps$ as
\bse
\be
\label{e:perturbation_expansion-a}
r_n(t) &=& \sum_{k=0}^\iy \eps^k r_{kn}(t,\tau),\\
f_n(t) &=& \sum_{k=0}^\iy \eps^k f_{kn}(t,\tau),
\label{e:perturbation_expansion-b}
\ee
\ese
where we explicitly separate the dependence on the two timescales. Then

\benn
\ddot r_{kn} (t,\tau) &=& \p_t^2 r_{kn} + 2 \eps \p_t \p_\tau r_{kn} + \eps^2 \p_\tau^2 r_{kn},\\
\ddot f_{kn} (t,\tau) &=& \p_t^2 f_{kn} + 2 \eps \p_t \p_\tau f_{kn} + \eps^2 \p_\tau^2 f_{kn}.
\eenn

At zeroth order in $\eps$, with $D = 0$,  Eq.~\eqref{e:rho_second_order} becomes
\benn
\sum_{n=1}^{V} \p_t^2 r_{0n} \gvr_{vn}=-\sum_{n=1}^{V} \gl_n^2 r_{0n} \gvr_{vn}.
\eenn
The modes $\gvr_{vn}$ are orthonormal, so the terms decouple into separate harmonic oscillators; $f_{kn}$ can be found from $r_{kn}$ using Eq.~\eqref{e:active_model_resc-a}. The leading order solution is then

\benn
r_{0n}(t) &=& A_{0n}(\tau) \cos(\gl_n t - \delta_n(\tau)), \\
f_{0n}(t) &=&-A_{0n} (\tau) \sin (\gl_n t - \delta_n(\tau)).
\eenn

At first order in $\eps$, with $g(\mu, \phi_e) = (\mu - \phi_e^2)$,
\benn
\sum_{n=1}^{V} (\p_t^2 r_{1n}+2 \p_t\p_\tau r_{0n})\gvr_{vn} =
 - \sum_{n=1}^{V} \gl_n^2 r_{1n}\gvr_{vn}
+ \sum_e \nabla_{ve}\left[\mu-\left(\displaystyle\sum_{n=1}^{E} f_{0n}\phi_{en} \right)^2\right]\sum_{\l=1}^{E} f_{0\l} \phi_{e\l}.
\eenn
Multiplying by $\gvr_{vm}$ and summing over $v$, we find
\be
\p_t^2 r_{1m}+2 \p_t\p_\tau r_{0m} = 
 -\gl_m^2 r_{1m} + \gl_m \left[\mu f_{0m} - \displaystyle \sum_e \phi_{em} \left(\sum_{n=1}^{E} f_{0n}\phi_{en} \right)^3 \right]. 
 \label{e:rho_order_eps}
\ee

\section{Leading order amplitude dynamics}
\label{s:amplitude_dynamics}
In order for the expansion in Eqs.~\eqref{e:perturbation_expansion-a} and \eqref{e:perturbation_expansion-b} to make sense, the magnitudes of the summands $r_{kn}$ and $f_{kn}$ must remain bounded. From Eq.~\eqref{e:rho_order_eps}, $r_{1m}$ is a harmonic oscillator with natural frequency $\lambda_m$ driven by the zeroth order oscillations. It will have bounded oscillations only if the resonant terms in Eq.~\eqref{e:rho_order_eps}, those that drive $r_{1m}$ at its natural frequency, are zero. Finding the resonant terms and setting them to zero will fix the leading order mode amplitudes $A_n(\tau)$.

Expanding the cube in Eq.~\eqref{e:rho_order_eps} gives
\begin{align}
\p_t^2 r_{1m}+2 \p_t\p_\tau r_{0m} =& 
 -\gl_m^2 r_{1m} + \gl_m \bigg[\mu f_{0m} -  \sum_e \phi_{em}\sum_{k,\ell,n=1}^{E} f_{0k}\phi_{ek}f_{0n}\phi_{e\ell}f_{0n}\phi_{en}\bigg]  \nonumber \\
 =&-\gl_m^2 r_{1m} + \gl_m \bigg[\mu f_{0m} +\sum_{k,\ell,n=1}^{E}\left(\sum_e \phi_{em} \phi_{ek}\phi_{e\ell}\phi_{en}\right) \nonumber
 \\
 & 
\hspace{4cm} \times A_{0k}A_{0\ell}  A_{0n}\sin (\gl_k t - \delta_k)\sin (\gl_\ell t - \delta_\ell)\sin (\gl_n t - \delta_n)
 \bigg]. \label{e:order_eps_expanded}
\end{align}
Now, the product of sines can be expanded into
\begin{align*}
\sin (\lambda_k t-\delta_k) \sin (\lambda_\ell  t-\delta_\ell) \sin (\lambda_n t-\delta_n)
&=
\frac{1}{4} \Big[\sin (\delta_k-\delta_\ell-\delta_n-\lambda_k t+\lambda_n t+\lambda_\ell  t)\nonumber\\
&\qquad -\sin (\delta_k-\delta_\ell+\delta_n-\lambda_k t-\lambda_n t+\lambda_\ell  t)\nonumber\\
&\qquad-\sin (\delta_k+\delta_\ell-\delta_n-\lambda_k t+\lambda_n t-\lambda_\ell  t)\nonumber\\
&\qquad +\sin (\delta_k+\delta_\ell+\delta_n-\lambda_k t-\lambda_n t-\lambda_\ell  t)\Big].
\end{align*}
We seek only resonant terms, which only occur when $\pm\gl_k$, $\pm\gl_\ell$, and $\pm\gl_n$ sum to $\gl_m$. This happens most often in one of two ways. First, we might have $k = \ell$ and $n = m$ or similar. Alternatively, we might have degenerate modes, $\gl_k = \gl_\ell$ and $\gl_n = \gl_m$. However, we ignore the latter possibility because degeneracies add significant analytic complications, including nontrivial dynamics of their relative phases. We also ignore the rare possibility of resonant terms arising from interactions of modes with three or four distinct singular values. The results we get with these assumptions closely match simulated time series (Fig.~3e-g), suggesting that the existence of degeneracies has little impact on the dynamics of nondegenerate modes.

The remaining resonant terms in Eq.~\eqref{e:order_eps_expanded} must cancel so that $r_{1m}$ is not an oscillator of frequency $\gl_m$ driven at frequency $\gl_m$. Thus,  
\begin{align*}
2 \p_t\p_\tau r_{0m} 
 = \gl_m \left[\mu f_{0m} +\frac{1}{4} \left(\sum_e \phi_{em}^4\right)
 A_{0m}^3 (3 \sin(\gl_m t - \gd_m))+
  3 \!\! \sum_{k=1,k\neq m}^{E}\left(\sum_e \phi_{em}^2 \phi_{ek}^2\right)
A_{0k}^2  A_{0m} \frac{1}{4} \left(2\sin(\gl_m t- \gd_m) \right)
 \right]\! .
\end{align*}
Substituting in $r_{0m}$ and $f_{0m}$,
\begin{align*}
-2 A_{0m}' \gl_m \sin(\gl_m t - \gd_m) +2 \gl_m^2 \cos(\gl_m t -\gd_m) \gd_m'
 = \\
 \gl_m \left[-\mu A_{0m} \sin(\gl_m t - \gd_m) +\frac{1}{4}\left(\sum_e \phi_{em}^4\right)A_{0m}^3 (3 \sin(\gl_m t - \gd_m))+\right.&\left.
  3\!\! \sum_{k=1, k\neq m}^{E}\left(\sum_e \phi_{em}^2 \phi_{ek}^2\right)
A_{0k}^2  A_{0m} \frac{1}{4} \left(2\sin(\gl_m t- \gd_m) \right)
 \right]\! ,
\end{align*}
where primes denote differentiation with respect to $\tau$.
For this to hold for all $t$ we need the coefficients of the sine and cosine terms to separately cancel. From the cosine term, $\gd_m'= 0$; from the sine term,

\begin{align*}
A_{0m}'=  \frac{1}{2}A_{0m}\left( \mu - \frac{3}{4} \left(\sum_e \phi_{em}^4\right)A_{0m}^2  -
  \frac{3}{2}\sum_{k=1, k\neq m}^{E}\left(\sum_e \phi_{em}^2 \phi_{ek}^2\right)
A_{0k}^2  \right)
\equiv \f 1 2 A_{0m} \left(\mu - \sum_{k=1}^E P_{mk} A_{0k}^2\right),
\end{align*}
where the matrix $\matx{P}$ has entries $P_{mk} = \tfrac{3}{2} (1 - \tfrac{1}{2}\gd_{mk})\sum_e \phi_{em}^2 \phi_{ek}^2$. Rewriting in terms of the squared amplitudes,
\be
\frac{d}{d\tau} (A_{0m}^2) = 2 A_{0m} A_{0m}' = A_{0m}^2 \left(\mu - \sum_{k=1}^E P_{mk} A_{0k}^2\right).
\label{e:amplitude_dynamics}
\ee
As a matrix equation, with $x_m = A_{0m}^2$, this reads
\be
\mbf x' = \mbf x \odot (\mu \mbf 1 - \matx{P} \mbf x),
\label{e:xprime}
\ee
where $\mbf 1$ denotes the vector of ones and $\odot$ is the component-wise product.

To find stationary points, we  set $ \mbf x \odot (\mu \mbf 1- \matx{P} \mbf x) = 0$.
The obvious way to solve Eq.~\eqref{e:xprime} for all stationary points is to exhaustively search over combinations of active modes: on picking certain elements of $\mbf x$ to be zero,  the remaining nonzero entries $\hat{\mbf x}$ are found by solving $\hat{\matx{P}} \hat{\mbf  x} =\mu \mbf 1$,
where $\hat{\matx{P}}$ is $\matx{P}$ restricted to those modes chosen to be nonzero. 
Stability of a fixed point $\mbf x_0$ then follows by standard perturbation analysis: inserting a small perturbation $\mbf x_0 + \gd \mbf x(\tau)$ into Eq.~\eqref{e:xprime} gives
\begin{align*}
\gd \mbf x' 
= \gd \mbf x - \mbf x_0 \odot (\matx{P} \gd \mbf x)  - (\matx{P} \mbf x_0) \odot \gd \mbf x + O(\gd\mbf  x^2) 
\equiv \matx{M} \gd \mbf x + O(\gd \mbf x^2),
\end{align*}
where $\matx{I}$ denotes the identity matrix,
and the eigenvalues of $\matx{M}$ then determine stability in the usual fashion.

\section{Accuracy of Rayleigh friction approximation}

To verify that the Rayleigh friction approximation does not significantly impact the results, we check the amplitude and stability of single modes for the full model with $g(\mu, \phi_e) = (\mu - \phi_e^2)/(1 + \phi_e^2)$ on all edges. Here setting the first order secular terms to zero in a perturbation expansion with $A_{0n} = A_{0p} \gd_{np}$ leads to
\be
\label{e:A_p_mu}
A_{0p}^2=(\mu+1) \sum_e\left( 2 - 2\sqrt{\f 1 {1+A_{0p}^2 \gf_{ep}^2}} \right).
\ee
Numerically solving Eq.~\eqref{e:A_p_mu} for $\mu = 1$ yields solutions within a few percent of the Rayleigh approximation solution $ 1/ {\sqrt{P_{pp}}}$ which additionally match numerical simulations of the full model even for $\eps$ as large as $0.5$ (Fig.~S\ref{f:amplitude-mu}).

%%%%%%%%%%%%%
\begin{figure*}[t]
\includegraphics[width=2.5in]{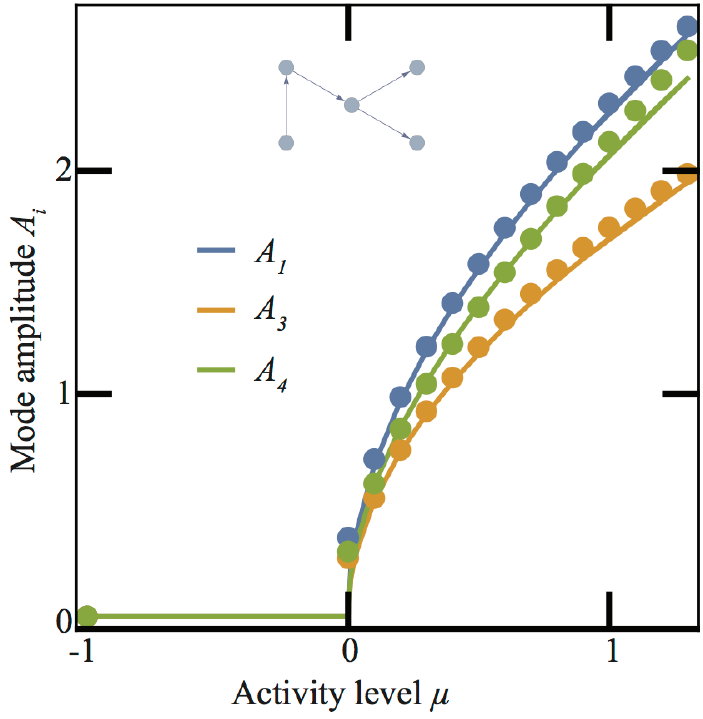}
\caption{Steady state amplitudes $A_i$ as a function of activity $\mu$ for the tree pictured undergo a Hopf bifurcation as $\mu$ crosses~0. Dots are long-time root-mean-square amplitudes from simulations started in each mode; lines are numerical solutions of Eq.~\eqref{e:A_p_mu}.  Mode $A_2$ is too unstable to reliably observe in simulations, so it is omitted. For $\mu < 0$, all amplitudes go to zero in simulations; the dot included in that region is at $\mu = -1$ where the friction is purely passive. Some deviations between simulation and analytics are expected because the simulations do not use the Rayleigh friction approximation and $\eps \neq 0$. Parameters were $\eps = 0.5$ and $D = 0$.
\label{f:amplitude-mu}}
\end{figure*}
%%%%%%%%%%%%%

When the system transitions from no energy input to active flow, the steady state amplitudes will grow with $\mu$. If we assume $\mu \ll 1$ (so $A_p \ll 1$) and expand the square root to order $A_p^4$, we find
$
A_p^2 + O(A_p^4) = {\mu} /{ P_{pp}},
$
exactly matching the Rayleigh friction result. The scaling $A_p \sim \sqrt{\mu}$ is typical of a supercritical Hopf bifurcation.

\section{Attractor characteristics on tree networks}

%%%%%%%%%%%%%%%%%%%%%%%%%%%
\begin{figure*}[b]
\includegraphics[width=0.9\linewidth,]{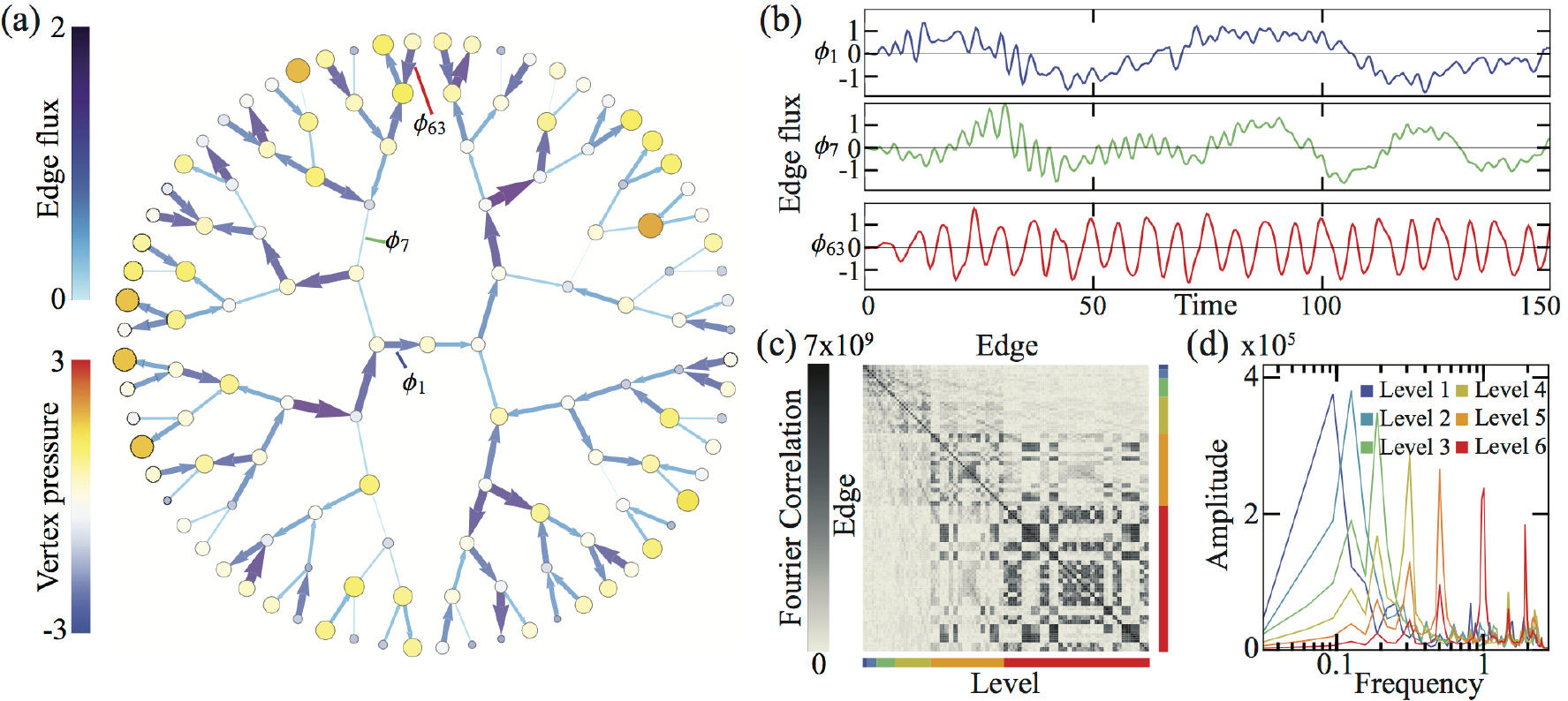}
\includegraphics[width=0.9\linewidth,]{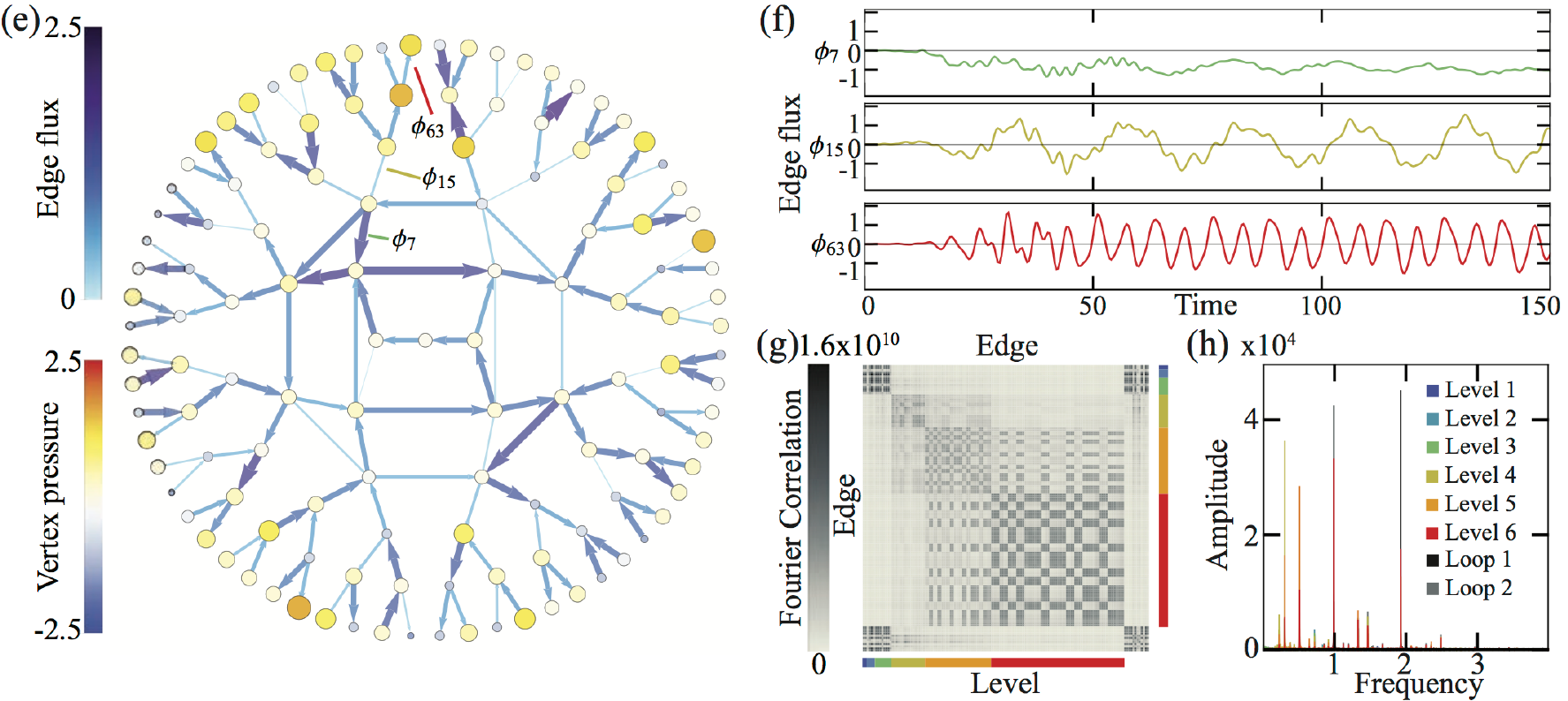}
\caption{
Activity causes depth-dependent separation of time scales on a large tree. 
(a) Most pressure variation occurs near the leaves on large binary trees (Movie 4). 
(b) The tree in (a) develops an activity-driven steady state with slow oscillations in the center and fast oscillations near the edges, as illustrated by the flux $\phi_e$ on the three edges labelled in (a).
(c) Unnormalized correlations between the Fourier transforms of the flux through the edges of the tree in (a), with phases ignored. Colors indicate the tree level of the tail vertex of the edge. There are strong correlations within each level and between neighboring levels, but low correlations for edges in widely-separated levels. 
(d) Frequency spectra of each tree level, computed by taking Fourier transforms of the edge fluxes as in (c) and averaging the magnitudes across all edges at each level. A distinct primary oscillation frequency for each level can be seen, which increases with distance from the tree center. Simulation parameters in all panels are $\eps = 0.5$, $\mu = 1$, and $D = 10^{-3}$.
(e-h) While adding edges in the center leads to steady flow on cycles there, frequency still increases with distance from the center in the outer, tree-like sections.
}
\label{f:large_tree}
\end{figure*}
%%%%%%%%%%%%%%%%%%%%%%%%%%%

The mode interactions of Eq.~\eqref{e:xprime} can lead to complex oscillation patterns dependent on global, not local, topology, as shown for a 127-vertex complete binary tree in Movie~4 and Fig.~S\ref{f:large_tree}.
After initializing with zero pressure variation and flux, the system settles into quasi-steady states with dramatically different dynamics in separate regions of the tree (Fig.~\ref{f:large_tree}a,b). Flux in edges near the leaves of the tree tends to oscillate rapidly, driving large pressure fluctuations in nearby vertices, whereas flux oscillations near the root  are comparatively slow with nearly constant pressure in the vertices (Fig.~\ref{f:large_tree}b,d). Since, apart from the root and leaves, each vertex has the same local topology, the different time scales emerge from the interaction of the local active friction with the global structure of the tree.

%%%%%%%%%%%%%
\begin{figure*}[h]
\includegraphics[width=3.5in]{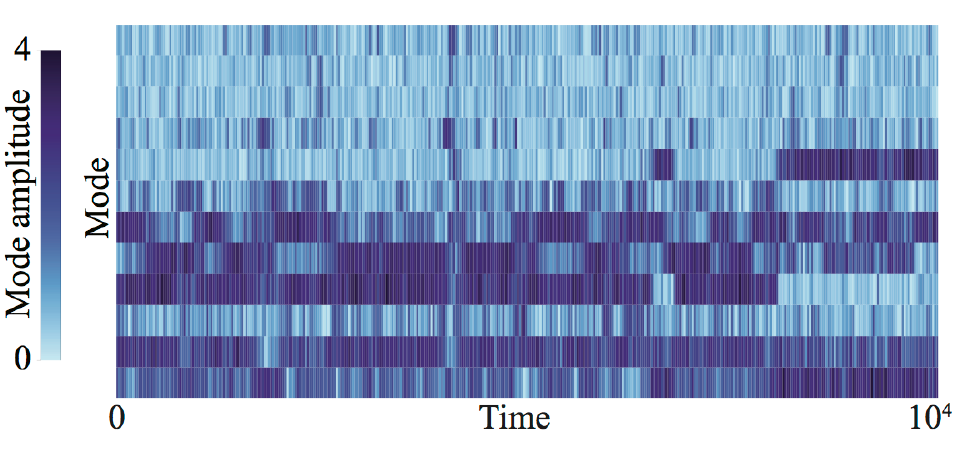}
\caption{Lower energy modes transition more often for the graph in Fig.~3e of the Main Text.  Modes are ordered by frequency from high (top) to low (bottom). Simulation parameters are $\eps = 0.5$, $\mu = 1$, $D = 5 \times 10^{-3}$, identical to those in Fig. 3. Note that rows 7 and 8, the two modes that switch on and off most, are degenerate.
\label{f:attractor_types}}
\end{figure*}
%%%%%%%%%%%%%

A comprehensive and precise characterization of the relative lifetimes of different attractors in large active flow networks remains out of reach with current numerical methods, in part because the range of noise levels low enough to observe state selection and high enough to observe transitions is quite small. Such a fine-tuning between thermal and active transport processes is a characteristic feature of many, if not all,   biological systems that function optimally in a narrow temperature range: bacterial flagellar motors are designed to barely beat Brownian diffusion at  room temperature, ATP-driven intracellular transport is tuned such that it improves moderately over thermal diffusion, and so-on.  Another well-known example in this context is  stochastic resonance in driven multistable systems~\cite{1998GaHaJuMa}. However, as all these systems typically exhibit exponential Arrhenius-type waiting times, it is practically impossible to completely explore their attractor statistics in the moderate-to-weak noise regime, except for the simplest two-state systems~\cite{Hanggi1990_RevModPhys}.

\par
Nevertheless, long simulation runs as shown in Fig.~S\ref{f:attractor_types} offer some insight into the qualitative behavior of attractors in active flow networks. Specifically, our simulations suggest that, while there is considerable variation in the relative occupancy of different attractors, stable states can be approximately divided in two classes: (1) states with one high energy mode at high amplitude and a few low energy modes at low amplitude and (2) states with multiple low-energy modes active at moderate amplitude, some of them degenerate. States of type (2) tend to quickly transition to other states of type (2) (Fig.~S\ref{f:attractor_types}); states of type (1) have a wide range of lifetimes but no obvious transition patterns.

\section{Networks with cycles}
We focus on tree networks in this paper as they allow substantial analytical progress. However, Eqs.~\eqref{e:active_model_resc} can be applied without modification to networks with cycles. Cycles correspond to right singular vectors ${\bf \phi}_n$ of $\nabla^\top$ with singular value zero. As these are always degenerate, we expect the conclusions of Section~\ref{s:amplitude_dynamics} to be most accurate when there are few or no cycles. Alternatively, on a weighted graph where the edges of high conductance form a tree, the attractor characteristics will be similar to the attractors on that tree (Fig.~1; all modes pictured in Fig.~S\ref{f:figure1_extended}).

%%%%%%%%%%%%%
\begin{figure*}[b]
\includegraphics[width=6in]{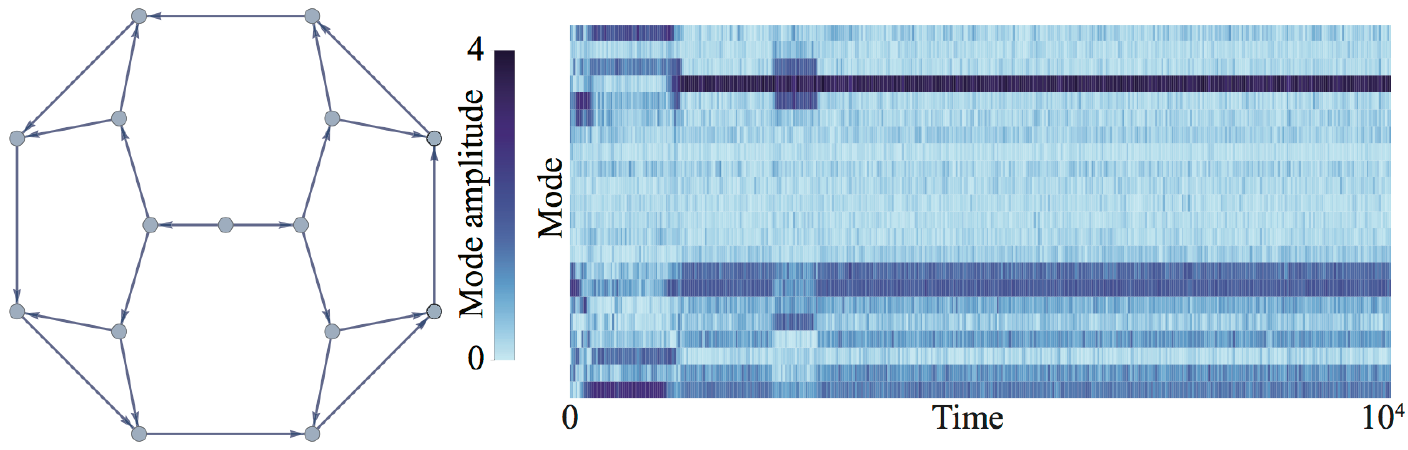}
\caption{States on graphs with cycles, like the one shown, tend to be more stable. Modes are ordered by frequency from high (top) to low (bottom). Note that the eight modes at the bottom, which are the only ones active in the lower half of the trace, are all cycles. Simulation parameters are $\eps = 0.5$, $\mu = 1$, $D = 5 \times 10^{-3}$.
\label{f:cycle_graph}}
\end{figure*}
%%%%%%%%%%%%%

Qualitatively, we find the same stochastic switching between states with subsets of modes active in simulations of Eqs.~\eqref{e:active_model_resc} on cyclic graphs even with equal weights, with the additional feature that cyclic modes are particularly stable and take longer to transition on average (Fig.~S\ref{f:cycle_graph}). For further discussion of similar dynamics on cycles, see \cite{Woodhouse2016}.

%%%%%%%%%%%%%
\begin{figure*}[t]
\includegraphics[height=7in]{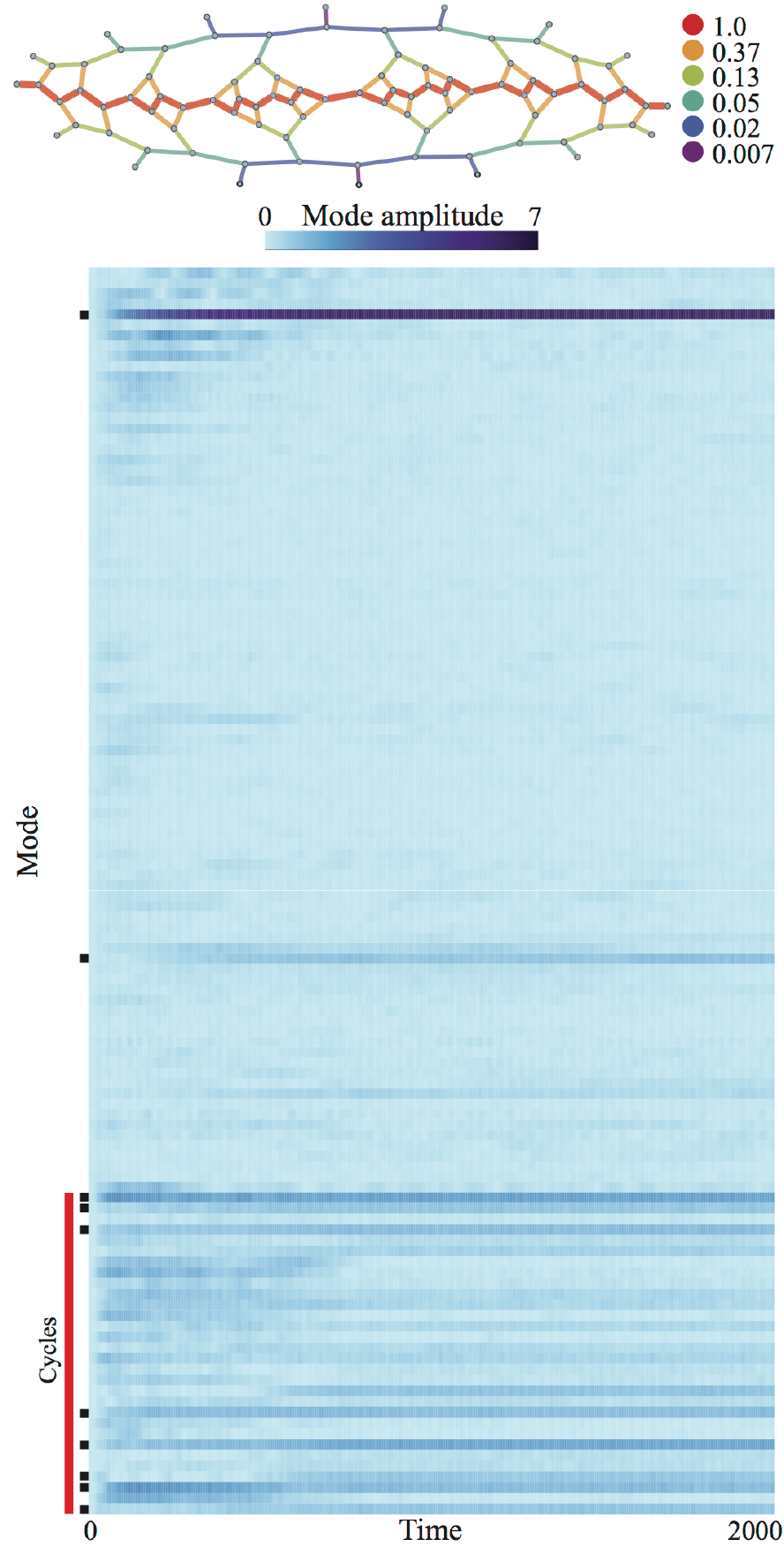}
\caption{
Including all of the modes from the simulation in Fig.~1 of the Main Text shows clear single mode selection on this weighted network. Edges a distance $d$ from the central red path were given weight $e^{-d}$.
Modes are ordered by frequency from high (top) to low (bottom); the last thirty modes, marked in red, are cycles. The modes pictured in Fig.~1 are marked in black.
\label{f:figure1_extended}}
\end{figure*}
%%%%%%%%%%%%%

\pagebreak
\section{Higher order oscillations}

Before, by setting resonant terms to zero, we found the slow dynamics of $A_n$. Now we look at the non-resonant terms driving $r_{1m}$ to find higher order effects. If we let
\benn
S_{i_1^{n_1} \ldots i_k^{n_k}} = \sum_e \prod _{j=1}^k \gf_{ei_j}^{n_j},
\eenn
assume the resonant terms are zero, and assume $A_{0m} = A_{0p} \gd_{mp} + A_{0q}\gd_{mq}$, the remainder of Eq.~\eqref{e:rho_order_eps} is 
\be
\p_t^2 r_{1m}+\gl^2_m r_{1m}
&=& \f 1 4 \gl_m \Big\{ S_{mp^3}A_{0p}^3 \sin(3\gl_p t)+3S_{mq^2p}A_{0p} A_{0q}^2\big[ \sin ((2\gl_q-\gl_p) t)-\sin ((2\gl_q +\gl_p) t) \big]
\nonumber  \\
&& \qquad\qquad +\, 3S_{mqp^2}A_{0p}^2 A_{0q}\big[ \sin ((2\gl_p-\gl_q) t)-\sin ((2\gl_p +\gl_q) t)\big] \nonumber\\
&& \qquad\qquad +\, S_{mq^3}A_{0q}^3 \sin(3\gl_q t) \Big\}.
\label{e:rho_order_eps_no_resonant}
\ee
Setting $m = p$ and only looking at the terms closest to resonance, we obtain
\benn
\p_t^2 r_{1p}+\gl^2_p r_{1p}
&\approx& \f 1 4 \gl_p \big\{ 3S_{q^2p^2}A_{0p} A_{0q}^2\sin ((2\gl_q-\gl_p) t)  
+ 3S_{p^3q}A_{0p}^2 A_{0q}\sin ((2\gl_p-\gl_q) t) \big\}.
\eenn
Thus
\benn
r_{1p} &\approx& c_1 \cos ((2\gl_q-\gl_p) t - \gd_1) + c_2 \cos ((2\gl_p-\gl_q) t - \gd_2),\\
f_{1p} &\approx& -c_1 \sin ((2\gl_q-\gl_p) t - \gd_1) - c_2 \sin ((2\gl_p-\gl_q) t- \gd_2),
\eenn
where
\benn
c_1 &=& \f {3}{4((2\gl_q-\gl_p)^2 - \gl_p^2)} \gl_p S_{q^2p^2}A_{0p}A_{0q}^2, \\
c_2 &=& \f {3}{4((2\gl_p-\gl_q)^2 - \gl_p^2)} \gl_p S_{qp^3}A_{0p}^2 A_{0q} .
\eenn
The energy in this mode to first order in $\eps$ is
\begin{align*}
H_p &= \f {\gl_p^2}{2} \left((r_{0p}^2 + \eps r_{1p})^2 + (f_{0p}^2 + \eps f_{1p})^2\right) + O(\eps^2)
\\
& = \f {\gl_p^2}{2} \Big\{ A_{0p}^2 + 2\eps A_{0p} \big[ c_1 \cos((2\gl_q - 2\gl_p) t) +c_2\cos((\gl_p-\gl_q)t)\big] \Big\} + O(\eps^2),
\end{align*}
exhibiting an order $\eps$ time dependence. The coefficients $c_1$ and $c_2$ are small unless $\gl_p \approx \gl_q$. If we kept the frequency $3\gl_p$, $3\gl_q$, $2\gl_p + \gl_q$, and $2\gl_q + \gl_p$ terms from Eq.~\eqref{e:rho_order_eps_no_resonant}, we would find energy oscillations with frequencies $2\gl_p$, $2\gl_q$, $3\gl_q - \gl_p$, and $\gl_p +\gl_q$ (Fig.~S\ref{f:energy_oscillations}); those oscillations have smaller amplitudes as the driving is farther from resonance.

\begin{figure*}[t]
\includegraphics[width= \linewidth]{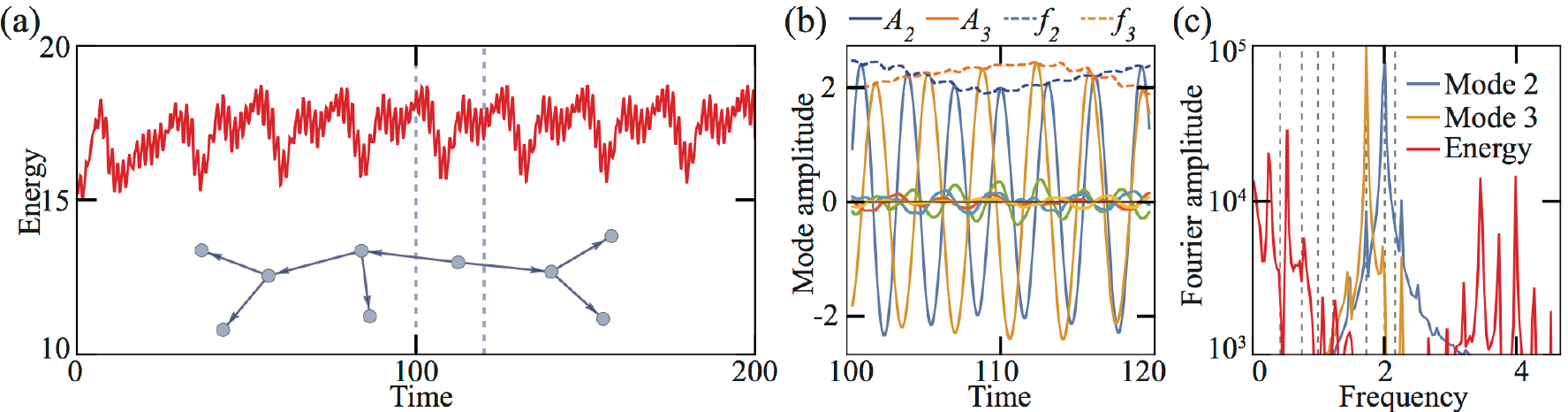}
\caption{Slow global oscillations emerge from the fast active dynamics.
(a) First order considerations fix a constant mean flow energy; higher order effects cause significant slow oscillations about that mean.
Simulation parameters were $\mu = 1$, $\eps = 0.5$, and $D=0$; the tree used is inset.
(b)  The mode amplitudes $A_2$ and $A_3$, like the energy, oscillate much more slowly than the harmonic oscillations of $f_2$ and $f_3$. All other mode amplitudes (unlabelled traces) are close to zero.
(c) Frequency spectra of the two active modes and the energy $H$ for the simulation in (a) and (b). The energy oscillates due to higher-order interactions between modes at frequencies that are linear combinations of active mode frequencies, not the harmonic frequencies alone (dashed lines). 
}
\label{f:energy_oscillations}
\end{figure*}

\section{Noise and thermalization}

In Eqs.~\eqref{e:active_model_resc-a} and \eqref{e:active_model_resc-b} we add Gaussian white noise only to the flux as a physically intuitive source of random fluctuations that preserve mass conservation. However, even with purely passive friction, this does not lead to equipartition of energy as seen in thermal systems. 

Written as stochastic differential equations with $g(\mu, \phi) = -1$, Eqs.~\eqref{e:active_model_resc-a} and \eqref{e:active_model_resc-b} become
\bse
\label{e:active_model_noise}
\be
\label{e:active_model_noise-a}
d\gvr_v &=&  \sum_e\nabla_{ve} \phi_e dt,
\\
\label{e:active_model_noise-b}
d\phi_e &=&-  \sum_v \nabla^\top_{ev}\gvr_v dt 
 -\eps\phi_e dt + \sqrt{2 D} d\tilde B_e(t),
\ee
\ese
where each $\tilde B_e(t)$ is standard Brownian motion. 
The components of the $E$-dimensional Brownian motion ${\bf B}(t) = (\tilde B_1, \ldots, \tilde B_E(t))$ in any orthonormal basis are also standard Brownian motions, so we can rewrite the system in the mode basis as
\bse
\label{e:passive_model_noise_modes}
\be
\label{e:passive_model_noise_modes-a}
dr_n &=&  \gl_n f_n dt,
\\
\label{e:passive_model_noise_modes-b}
df_n &=&-  \gl_n r_n dt 
 - \eps f_n dt + \sqrt{2 D}  d B_n(t).
\ee
\ese
The associated Fokker-Planck equation for the probability distribution $p({\bf r}, {\bf f}, t)$ is
\be
\p_t p =\sum_n\left[ -\f {\p }{\p r_n} (\gl_n f_n p) + \f {\p}{\p f_n} ( \gl_n r_n p) +\f \p {\p f_n}(\eps f_n p)+ D \f{\p^2p}{\p f_n^2}\right]
\nonumber 
\ee
with $p\rightarrow 0$ as $r_n,f_n \rightarrow \infty$ and $p$ integrating to 1.
Now, without friction or noise, the dynamics are governed by the Hamiltonian
\benn
H = \f 1 2 \gvr_v \nabla_{ve} \nabla_{eu} \gvr_u + \f 1 2 \phi_e \nabla_{ev} \nabla_{va} \phi_a 
 =\f 1 2 \sum_n \gl_n^2 (r_n^2 + f_n^2) \equiv \sum_n H_n.
\eenn
If $p$ is a function of the $H_n$ alone, the Fokker-Planck equation in steady state reduces to
\benn
0 =\sum_n\f \p {\p f_n}(\eps f_n p)+ D \f{\p^2p}{\p f_n^2},
\eenn
which has solution
\benn
p(H_1,\ldots, H_M) \propto \prod_{n=1}^M e^{-\f {H_n}{kT_n}},
\eenn
where $
kT_n =  {\gl_n^2 D} / {\eps}
$.

Loosely, adding noise this way couples each mode to a heat bath with a distinct temperature. The result is equipartition of amplitude, not energy: the long-time average $\langle A_n^2\rangle$ is independent of $n$. Adding weak coupling between modes by making $\mu > -1$ does not change this. 

To get equipartition of energy one could change the coupling to noise, replacing the final term in Eq.~\eqref{e:passive_model_noise_modes-b} with
\benn
\sqrt{2 D_n}  d B_n(t) \equiv \f {\sqrt{2 D}} {\gl_n}  d B_n(t).
\eenn
This is only possible for $\gl_n \neq 0$, which precludes cyclic modes.
Equation~\eqref{e:active_model_resc-b} becomes
\benn
d\phi_e &=& - \sum_v \nabla^\top_{ev}\gvr_v dt 
 + \eps g(\mu, \phi_e) \phi_e dt + \sum_n \f 1 {\gl_n}\phi_{en} \sqrt{2 D}  d B_n(t).
\eenn
The previous analysis goes through identically, leading to
$
kT_n =  {\gl_n^2 D_n} /{\eps} =  D /\eps 
$.

\section{Differential growth rates}

While the $E/4$ active modes per state that we observe is significantly reduced relative to the total number of modes available, it is still a not insignificant fraction of $E$. There are, however, several straightforward generalizations of our model that may lead to more strict mode selection.
We discuss two possibilities in this and the subsequent section: variations in activity across the network and variations in weights of vertices or edges.

For simplicity, we introduced
Eqs.~\eqref{e:active_model_resc} with a uniform activity level $\mu$ across the entire network. This leads to equal driving on all modes: if Eq.~\eqref{e:amplitude_dynamics} is initialized near zero, it can be linearized to 
\benn
\frac{d}{d\tau} (A_{0m}^2) = \mu A_{0m}^2,
\eenn
where all modes grow at the same rate. Mode selection occurs in this system only because of interactions between modes.

In many physical systems, however, differences in growth rate between modes are important for mode selection. For example, the Rayleigh-Plateau instability \cite{Rayleigh1878} causes fluid jets to break apart into droplets whose size is determined by the fastest growing unstable perturbation to the jet radius. Nonlinear mode competition akin to that in Eqs.~\eqref{e:active_model_resc} may only act on the subset of modes that grow quickly.

We can add this effect to our model by replacing $\mu$ in Eqs.~\eqref{e:active_model_resc} with edge-dependent parameters $\mu_e$. With the quadratic driving of Eq.~\eqref{e:quadratic_activity}, Eq.~\eqref{e:active_model_resc-b} becomes
\benn
\f {d{ \phi}_e}{dt} &=&-\sum_v \nabla^\top_{ev} \gvr_v + \eps \left(\mu_e -\phi_e^2 \right)\phi_e
+ \sqrt{2 D}  \xi_e(t)  .
\eenn
Following through the previous calculations with this change, Eq.~\eqref{e:rho_order_eps} becomes
\benn
\p_t^2 r_{1m}+2 \p_t\p_\tau r_{0m} = 
 -\gl_m^2 r_{1m} + \gl_m \left[\sum_{e, l} \phi_{em} \mu_e f_{0l}\phi_{el} - \displaystyle \sum_e \phi_{em} \left(\sum_{n=1}^{E} f_{0n}\phi_{en} \right)^3 \right]. 
\eenn
The first term inside the square brackets no longer simplifies, since the $\phi_{en}$ are not orthonormal with the weighting $\mu_e$. However, if we again ignore degeneracies, the only resonant term is $\sum_{e} \phi_{em}^2 \mu_e f_{0m}$
 from $l = m$. In this case, defining $\nu_m = \sum_e \phi_{em}^2 \mu_e$, Eq.~\eqref{e:amplitude_dynamics} then reads 
\be
\frac{d}{d\tau} (A_{0m}^2) = A_{0m}^2 \left(\nu_m - \sum_{k=1}^E P_{mk} A_{0k}^2\right),
\label{e:different_growth_rates}
\ee
where modes have distinct growth rates independent of their interactions. Alternatively, one could specify $\nu_m$ arbitrarily in Eq.~\eqref{e:different_growth_rates}, though this would require more complex changes in Eq.~\eqref{e:active_model_resc-b}  coupling activity across edges.

\begin{figure*}[b]
\includegraphics[width= \linewidth]{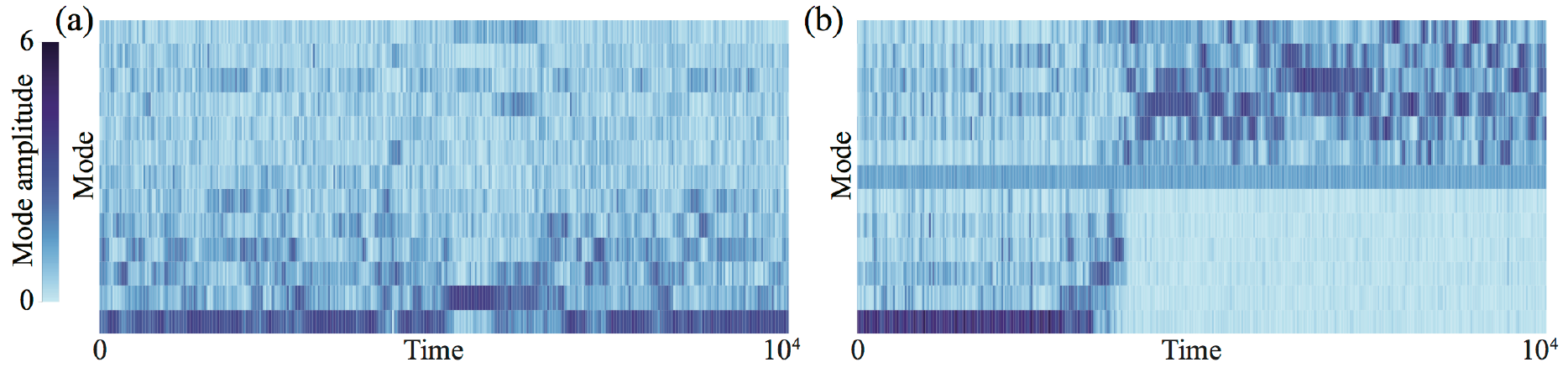}
\caption{The emergence of an activity-driven spectral band gap is exhibited by a simulation on a 14-vertex path with (a) all weights equal to 1 and (b) alternating vertex weights $1$ and $5$. Modes are ordered by frequency from high (top) to low (bottom). Note that in (b) the central $n=7$ mode is always active and the low energy states on the right half of the plot are significantly more suppressed than they ever are in (a). The qualitative difference is due to the presence of vertices with unequal weights, not the overall scale of the vertex weights; changing vertex weights uniformly is equivalent to rescaling other parameters. Parameters were $\mu = 1.2$, $D = 5\times 10^{-3}$, and $\eps = 0.5$. Both simulations used the same random seed.
}
\label{f:alternating_weights}
\end{figure*}

\section{Band gaps}

In addition to distinct activity levels $\mu_e$ across edges, we can also introduce edge weights $\gamma_e$ or vertex weights $m_v$ that vary across the network. Changing the conductances $\gamma_e$ and volumes $m_v$ changes our system in two ways: first, by changing the modes to the singular vectors of $\gamma^*_{ve}$; and second, by changing the coupling matrix to $\tilde P_{mk} = \tfrac{3}{2} (1 - \tfrac{1}{2}\gd_{mk})\sum_e \gamma_e^{-1}\phi_{em}^2 \phi_{ek}^2$, which depends explicitly on the edge weights. 

Such changes are known to cause qualitative changes in the physics of classical spring-mass networks, including the introduction of band gaps.
In an infinite one-dimensional line of beads of equal mass $m$ connected by springs with equal spring constant $f$, for example,  the dispersion relation between frequency $\omega$ and wavenumber $q$  is
\benn
\omega(q) = 2\sqrt{\f {f}{m}} \left| \sin\left( \f{qa}{2}\right)\right|,
\eenn
where $a$ is the size of the unit cell, in this case equal to distance between adjacent beads \cite{Misra2011}. If instead of equal masses the beads alternate between a smaller mass $m_1$ and larger mass $m_2$, the dispersion relation splits into two branches,
\benn
\omega(q)^2_\pm = f\left(\f{1}{m_1} + \f{1}{m_2}\right) \pm f \sqrt{ \left(\f{1}{m_1} + \f{1}{m_2}\right)^2 - \f{4}{m_1m_2} \sin^2\left(\frac{qa}{2}\right)}.
\eenn
Here a unit cell has two beads, so the distance between beads is $a/2$. At $q= \pi/a$, there is a gap between $\omega_+ = \sqrt{2f/m_1}$ and $\omega_- = \sqrt{2f/m_2}$. This band gap shows up in a finite system as a large difference in frequency between modes above and below the gap.

Since varying what are effectively vertex weights causes such a clear qualitative change in behavior in the spring system, we can reasonably expect similar changes in our model.
Simulations on paths with alternating vertex weights show a distinct separation of of low- and high-energy states not present with uniform weights (Fig.~S\ref{f:alternating_weights}), with stronger and more consistent suppression of the low-energy states and few transitions across the band gap created by nonuniform weights. Band gaps in more realistic topologies may have similar effects, allowing for enhanced control of the large-scale behavior.

%\bibliography{active_networks_references}

\end{widetext}

\end{document}